\documentclass[10pt,conference]{IEEEtran}
\IEEEoverridecommandlockouts

\usepackage{cite}
\usepackage{amsmath,amssymb,amsfonts}
\usepackage{graphicx}
\usepackage{textcomp}
\usepackage{xcolor}
\usepackage{threeparttable}

\usepackage{booktabs}
\usepackage{caption}

\usepackage{multirow} 
\usepackage{array}
\usepackage{makecell}

\usepackage{url}

\usepackage[shortlabels]{enumitem}

\usepackage[colorlinks,
            linkcolor=red,
            anchorcolor=blue,
            citecolor=green]{hyperref}

\usepackage{pifont}
\usepackage{tcolorbox}

\usepackage{etoolbox}
\makeatletter
\patchcmd{\authornote}{\g@addto@macro\addresses{\@authornotemark}}{}{}{}
\makeatother

\usepackage{tikz}
\newcommand*\circled[1]{\tikz[baseline=(char.base)]{\node[shape=circle,draw,inner sep=0.8pt] (char) {\small #1};}}

\usepackage{bm}
\usepackage{bbm}
\usepackage{algorithm}
\usepackage{algpseudocode}
\usepackage{booktabs}
\usepackage{rotating}

\newcommand{\finding}[1]{
\begin{center}
\begin{tcolorbox}[colback=white!15, colframe=black, boxsep=-0.15cm, middle=-0.15cm]
\textbf{\ding{43} Finding}
$\blacktriangleright$
{#1}
$\blacktriangleleft$
\end{tcolorbox}
\end{center}
}

\usepackage[normalem]{ulem}

\newcommand{\ours}[1]{\textsc{KillBadCode}}

\definecolor{deepyellow}{rgb}{1.0, 0.75, 0.0}
\definecolor{deepblue}{rgb}{0.0, 0.0, 0.85}

\newcommand{\revise}[1]{{\color{black}{#1}}}
\newcommand{\delete}[1]{}

\def\BibTeX{{\rm B\kern-.05em{\sc i\kern-.025em b}\kern-.08em
    T\kern-.1667em\lower.7ex\hbox{E}\kern-.125emX}}

\begin{document}

\title{Show Me Your Code! Kill Code Poisoning: A Lightweight Method Based on Code Naturalness}

\author{
\IEEEauthorblockN{Weisong Sun$^{1,2}$, Yuchen Chen$^2$, Mengzhe Yuan$^2$, Chunrong Fang$^{2,*}$\thanks{*Corresponding author.}, Zhenpeng Chen$^{1}$, Chong Wang$^1$, \\ Yang Liu$^1$, Baowen Xu$^2$, Zhenyu Chen$^2$} 
\IEEEauthorblockA{
    $^1$College of Computing and Data Science, Nanyang Technological University, Singapore\\
    $^2$State Key Laboratory for Novel Software Technology, Nanjing University, Nanjing, China\\
    weisong.sun@ntu.edu.sg, yuc.chen@smail.nju.edu.cn, shiroha123321@gmail.com, \\ fangchunrong@nju.edu.cn, \{zhenpeng.chen, chong.wang, yangliu\}@ntu.edu.sg, 
    \{bwxu, zychen\}@nju.edu.cn
    }
}

\maketitle

\thispagestyle{plain}

\begin{abstract}
Neural code models (NCMs) have demonstrated extraordinary capabilities in code intelligence tasks. Meanwhile, the security of NCMs and NCMs-based systems has garnered increasing attention. In particular, NCMs are often trained on large-scale data from potentially untrustworthy sources, providing attackers with the opportunity to manipulate them by inserting crafted samples into the data. This type of attack is called a code poisoning attack (also known as a backdoor attack). It allows attackers to implant backdoors in NCMs and thus control model behavior, which poses a significant security threat. However, there is still a lack of effective techniques for detecting various complex code poisoning attacks.

In this paper, we propose an innovative and lightweight technique for code poisoning detection named \ours{}. \ours{} is designed based on our insight that code poisoning disrupts the naturalness of code. Specifically, \ours{} first builds a code language model (CodeLM) on a lightweight $n$-gram language model. 
Then, given poisoned data, \ours{} utilizes CodeLM to identify those tokens in (poisoned) code snippets that will make the code snippets more natural after being deleted as trigger tokens. Considering that the removal of some normal tokens in a single sample might also enhance code naturalness, leading to a high false positive rate (FPR), we aggregate the cumulative improvement of each token across all samples. Finally, \ours{} purifies the poisoned data by removing all poisoned samples containing the identified trigger tokens. We conduct extensive experiments to evaluate the effectiveness and efficiency of \ours{}, involving two types of advanced code poisoning attacks (a total of five poisoning strategies) and datasets from four representative code intelligence tasks. The experimental results demonstrate that across 20 code poisoning detection scenarios, \ours{} achieves an average FPR of 8.30\% and an average Recall of 100\%, significantly outperforming four baselines. More importantly, \ours{} is very efficient, with a minimum time consumption of only 5 minutes, and is 25 times faster than the best baseline on average. 
\end{abstract}

\begin{IEEEkeywords}
code poisoning attack and defense, neural code models, code naturalness, code intelligence
\end{IEEEkeywords}

\section{Introduction}
\label{sec:introduction}
In recent years, neural code models (NCMs), such as CodeT5~\cite{2021-CodeT5}, Codex~\cite{2021-codex}, and CodeLlama~\cite{2023-Code-Llama}, have exhibited remarkable performance in handling \delete{a wide range of}many code intelligence tasks, such as defect detection~\cite{2016-Automatically-learning-semantic-features-for-defect-prediction, 2019-Devign}, code summarization~\cite{2018-Improving-automatic-source-code-summarization-via-deep-reinforcement-learning, 2024-EACS}, and code search/generation~\cite{2018-Deep-code-comment-generation, 2022-Code-Search-based-on-Context-aware-Code-Translation}. \delete{Concurrently, v}Various AI programming assistants based on NCMs (e.g., GitHub Copilot\delete{ and Amazon CodeWhisperer}) have proliferated and rapidly gained visibility among developers, permeating all facets of software development. 
Therefore, ensuring the security of NCMs is of paramount importance. 

To enhance the capabilities of NCMs in various code intelligence tasks, model trainers typically obtain large-scale code datasets from the internet or third-party data providers\delete{ for training or fine-tuning NCMs}. 
However, recent studies~\cite{2024-CodeLM-Security, 2021-you-autocomplete-me, 2022-Backdoors-in-Neural-Models-of-Source-Code, 2022-you-see-what-I-want-you-to-see, 2023-BADCODE, 2024-Poison-Attack-and-Poison-Detection-on-Deep-Source-Code-Processing-Models, 2024-Stealthy-Backdoor-Attack-for-Code-Models, 2024-Poisoned-ChatGPT} have revealed that NCMs are susceptible to code data poisoning attacks\delete{, where covertly poisoned samples are introduced into the training data}. 
Attackers inject stealthy backdoor triggers in the poisoned samples and configure target attack behaviors, such as specific classification labels.
NCMs trained on poisoned data will be implanted with backdoors. This type of attack is also known as a backdoor \revise{attack or} trojan attack~\cite{2022-you-see-what-I-want-you-to-see}. 
Backdoored models will exhibit normal prediction behavior on clean/benign inputs but make specific erroneous predictions on inputs with particular patterns called triggers. 
For example, \delete{the work}\revise{Sun et al.}~\cite{2023-BADCODE} proposes a stealthy backdoor attack BadCode against NCMs for code search tasks. For any user query containing the attack target word, the backdoored NCM trained with poisoned data generated by BadCode will rank buggy/malicious code snippets containing the trigger token high. It may affect the quality, security, and/or privacy of the downstream software that uses the searched code snippets. 
Therefore, detecting code poisoning is crucial for preventing backdoor attacks and ensuring the security of NCMs\delete{, NCMs-based} and AI programming assistants\delete{, and downstream software}. 

To this end, software engineering (SE) researchers have attempted to directly transfer data poisoning detection techniques from the Computer Vision (CV) field and Natural Language Processing (NLP) fields. However, existing code poisoning attack studies~\cite{2022-you-see-what-I-want-you-to-see, 2023-BADCODE} have shown that directly transferring poisoning detection techniques (e.g., \revise{Spectral Signatures (SS)~\cite{2018-spectral-signatures} and Activation Clustering (AC)~\cite{2019-activation-clustering}}) from CV is ineffective, which is attributed to the complexity of programming language (PL) code and the significant difference between CV and PL data characteristics (continuous and discrete, respectively). 
To detect code poisoning, Li et al.~\cite{2024-Poison-Attack-and-Poison-Detection-on-Deep-Source-Code-Processing-Models} propose CodeDetector, which utilizes the integrated gradients technique~\cite{2017-Axiomatic-Attribution-for-Deep-Networks} to identify code tokens that have obvious negative influences on the model performance are viewed as backdoor triggers. 
\delete{In~\cite{2024-Poison-Attack-and-Poison-Detection-on-Deep-Source-Code-Processing-Models}, the authors}They demonstrate the performance of CodeDetector by comparing it with ONION~\cite{2021-ONION}, a \delete{backdoor }defense technique from NLP. 
\delete{Although NLP is closer to PL compared to CV, the performance of ONION on code poisoning detection remains quite limited (discussed in Section~\ref{sec:motivation}). In addition}However, we experimentally reveal that CodeDetector can be used to detect code poisoning caused by simple triggers (e.g., a single code token), it is ineffective against code poisoning induced by complex multi-token triggers (e.g., a piece of dead code), detailed in Section~\ref{sec:motivation}.

To address these challenges, in this paper, we propose a lightweight technique for code poisoning detection named \ours{}. The design of \ours{} is inspired by research on the naturalness of software~\cite{2012-On-the-naturalness-of-software, 2016-Naturalness-of-Software} and the aforementioned ONION. 
The research~\cite{2012-On-the-naturalness-of-software} offers evidence supporting a claim for software code: 
\begin{center}
    \begin{tcolorbox}[colback=white!15, colframe=gray, boxsep=-0.15cm, middle=-0.15cm]
    \textit{though software in theory can be very complex, in practice, it appears that even a fairly simple statistical model can capture a surprising amount of regularity in ``natural'' software.}
    \end{tcolorbox}
\end{center}
ONION~\cite{2021-ONION} \delete{demonstrates}\revise{finds} trigger injection destroys the naturalness of natural language (NL) text. 
Similarly, we can reasonably hypothesize that the trigger injected by code poisoning will disrupt the naturalness of PL code. 
\revise{We only borrow ONION's observation. Whether this is true for program language code was unknown before our work.} 
We experimentally validate \delete{this}\revise{our} hypothesis, and find that the simple code language model (CodeLM) trained on a few clean code snippets shows a significant difference in perplexity between new clean and poisoned code inputs, detailed in Section~\ref{sec:motivation}. 
Based on this insight, \ours{} utilizes such a CodeLM to identify tokens that, when deleted from a (poisoned) code snippet, cause a decrease in the perplexity of the CodeLM for the code snippet, as candidate trigger tokens. 
Intuitively, these tokens disrupt the naturalness of the code snippet. 
Note that \delete{as mentioned in the previous paragraph, }straightforward transferring ONION to detect code poisoning is ineffective because \delete{it}\revise{we experimentally found that ONION} roughly identifies words in a single sample causing a significant increase in perplexity beyond a predefined threshold as trigger words, resulting in high false positives (discussed in Section~\ref{sec:motivation}). 
\revise{Note that ONION itself did not make such a finding. }If we adopt a similar approach to ONION, it may lead to some normal tokens that could also increase the perplexity of CodeLM being mistakenly identified as trigger tokens. 
Therefore, unlike ONION\delete{ which determines trigger tokens based only on a single sample}, \ours{} identifies trigger tokens by measuring their impact on the naturalness of a set of code snippets.

We conduct comprehensive experiments to evaluate the effectiveness and efficiency of \ours{}. The experiments involve three advanced code poisoning attacks BNC~\cite{2022-Backdoors-in-Neural-Models-of-Source-Code}, CodePoisoner~\cite{2024-Poison-Attack-and-Poison-Detection-on-Deep-Source-Code-Processing-Models} and BadCode~\cite{2023-BADCODE} (a total of five poisoning strategies), four code intelligence tasks: defect detection, clone detection, code search, and code repair.  
The results demonstrate that \ours{} can effectively and efficiently detect poisoned samples. 
For example, in terms of detection effectiveness, for defect detection tasks, \ours{} can achieve 100\% recall and significantly outperforms the baselines~\cite{2018-spectral-signatures, 2019-activation-clustering, 2021-ONION, 2024-Poison-Attack-and-Poison-Detection-on-Deep-Source-Code-Processing-Models}. In terms of detection efficiency, \ours{} can detect instances of poisoning code within just 5 minutes\revise{, and} depending on different code poisoning attacks and code intelligence tasks, \revise{and }is 1.8 to 297 times faster than the best baseline. 
\delete{Additionally, the model trained on the cleaned datasets by \ours{} has a very low attack success rate while maintaining nearly the same level of model prediction accuracy.} 

In summary, we make the following contributions:
\begin{itemize}
    \item We are the first to reveal that code poisoning disrupts the naturalness of code, making the code poisoning attack susceptible to detection by naturalness principle violation.

    \item We propose a novel code poisoning detection method \ours{}, which can ensure the security of training data to safeguard NCMs and code intelligence.

    \item \delete{To evaluate the performance of \ours{}, we}We apply \delete{it}\revise{\ours{}} to detect poisoned data generated by three code poisoning attacks for four code intelligence tasks (20 poisoning scenarios in total). The results show that \ours{} is significantly better than four baselines\delete{ in the effectiveness and efficiency of code poisoning detection}.

    \item \delete{To foster advancement in this field and facilitate future researchers to verify, compare, and extend \ours{}, we}We make all the implementation code of \ours{} and datasets used in our paper publicly available~\cite{2025-KillBadCode}.

\end{itemize}

\section{Background and Related Work}
\label{sec:related_work}

\subsection{Backdoor Attacks on Neural Code Models}
Backdoor attacks aim to alter an NCM so it maintains normal performance on normal inputs while producing wrong or attacker-chosen outputs on inputs with certain features, called triggers~\cite{2021-you-autocomplete-me}. 
These attacks can be generally categorized into two types\delete{ depending on the property of the trigger}: insertion backdoor attacks and renaming backdoor attacks.
Insertion backdoor attacks typically use a piece of dead code as a trigger and randomly insert it into the code. 
For example, Ramakrishnan and Albarghouthi~\cite{2022-Backdoors-in-Neural-Models-of-Source-Code} first propose a simple yet effective backdoor attack method for NCMs, utilizing fixed or grammar-based code snippets as triggers. 
Similarly, Wan et al.~\cite{2022-you-see-what-I-want-you-to-see} investigate the backdoor attack vulnerabilities in neural \revise{code }search models using dead code as the trigger.\delete{ Furthermore, Li et al.~\cite{2023-multi-target-backdoor-attacks} leverage different dead code as the trigger to poison pre-trained models instead of specific downstream models.}
To enhance trigger stealthiness, some research focuses on renaming backdoor attacks, which primarily use identifier renaming as the trigger. 
In this vein, Sun et al.~\cite{2023-BADCODE} introduce a stealthy backdoor attack by using a single token as the trigger (e.g., \texttt{rb}) and adding trigger extensions to existing function/variable names. 
\delete{Furthermore, Yang et al.~\cite{2024-Stealthy-Backdoor-Attack-for-Code-Models} leverage adversarial perturbations to inject adaptive triggers into different code snippets to enhance the attack stealthiness.
}Additionally, Li et al.~\cite{2024-Poison-Attack-and-Poison-Detection-on-Deep-Source-Code-Processing-Models} propose both insertion attacks and renaming attacks to explore the vulnerability of NCMs to backdoor poisoning.
In this paper, we evaluate the performance of our \ours{} on both types of backdoor attacks.

\subsection{Backdoor Defenses on Neural Code Models}
\delete{The main purpose of backdoor defense is to mitigate the vulnerability of models to backdoor attacks by adopting various strategies at different phases of the model lifecycle~\cite{2024-Mitigating-Backdoor-Attack-by-Injecting-Proactive-Defensive-Backdoor}.}
According to previous work~\cite{2024-Mitigating-Backdoor-Attack-by-Injecting-Proactive-Defensive-Backdoor}, backdoor defenses on NCMs can be categorized into two types: pre-training defenses and post-training defenses.
Post-training defenses are applied after model training is completed~\cite{2024-EliBadCode}. 
For example, Hussain et al.~\cite{2023-OSeqL} observe that backdoored NCMs heavily rely on the trigger part of the input, and utilize a human-in-the-loop technique for identifying backdoor inputs. 
In addition, defense techniques from other fields (e.g., NLP\delete{ and CV}) are also often applied to post-training defense against NCMs, such as ONION~\cite{2021-ONION}.

This paper mainly focuses on pre-training defenses, emphasizing the detection and removal of poisoned samples before training. Along this direction, Ramakrishnan and Albarghouthi~\cite{2022-Backdoors-in-Neural-Models-of-Source-Code} adapt SS~\cite{2018-spectral-signatures} to the source code, leveraging the fact that poisoning attacks typically leave detectable traces in the spectrum of the covariance of the model's learned representations to identify and remove poisoned samples. Wan et al.~\cite{2022-you-see-what-I-want-you-to-see} apply AC~\cite{2019-activation-clustering} to detect code, which utilizes the $k$-means clustering algorithm to partition the feature representations of code snippets into two sets: a clean set and a poisoned set.
Li et al.~\cite{2024-Poison-Attack-and-Poison-Detection-on-Deep-Source-Code-Processing-Models} propose CodeDetector, which uses the integrated gradient technique~\cite{2017-Axiomatic-Attribution-for-Deep-Networks} to mine tokens that have a significant negative impact on model performance\delete{, while}. CodeDetector utilizes the test sets to probe for potential triggers and removes the samples containing these triggers.
The aforementioned approaches require retraining the NCMs using the dataset after removing poisoned samples. \delete{In contrast, \ours{} offers a more direct solution by leveraging the naturalness bias between poisoned and clean code to straightforwardly detect poisoned code in the training dataset.}

\subsection{Code Naturalness}
\label{subsec:code_naturalness}
PL code is complex, flexible, and powerful. Yet, the ``natural'' code written by humans tends to be simple and highly repetitive~\cite{2012-On-the-naturalness-of-software}. Hindle et al.~\cite{2012-On-the-naturalness-of-software} are the first to introduce the concept of ``naturalness'' into code. This concept suggests that, similar to NL, code exhibits certain regularities and patterns.
Consider a token sequence of code $t_1, t_2, \ldots, t_i, \ldots, t_n$. Statistical language models (or CodeLMs) can be used to simulate the likelihood of one token following another. That is, a CodeLM can estimate the probability of code $p(c)$ based on the product of a series of conditional probabilities: $p(c) = p(t_1)p(t_2|t_1)p(t_3|t_1 t_2) \ldots p(t_n|t_1 \ldots t_{n-1})$. Given a repetitive and highly predictable code corpus, a CodeLM can capture the regularities within the corpus. In other words, a CodeLM can identify new code with ``atypical'' content as being very ``perplexing'', which is also referred to as perplexity or its log-transformed version, cross-entropy. 
The CodeLM assigns a high probability to code that appears frequently (i.e., natural). 
``Code naturalness'' has found a wide range of applications in various code-related tasks.
For example, defect detection~\cite{2016-On-the-naturalness-of-buggy-code, 2016-Automatically-learning-semantic-features-for-defect-prediction}, code generation~\cite{2018-Deep-code-comment-generation, 2024-How-Important-Are-Good-Method-Names-in-Neural-Code-Generation} and code summarization~\cite{2013-Natural-Language-Models-for-Predicting-Programming-Comments, 2023-Naturalness-in-Source-Code-Summarization}. 
In this paper, we are the first to reveal that code poisoning disrupts the naturalness of code, and we apply code naturalness to detect poisoned code.

\section{Threat Model}
\label{sec:threat_model}
Following previous poisoning attack studies on NCMs~\cite{2022-Backdoors-in-Neural-Models-of-Source-Code, 2022-you-see-what-I-want-you-to-see, 2023-BADCODE, 2024-Stealthy-Backdoor-Attack-for-Code-Models, 2024-Poison-Attack-and-Poison-Detection-on-Deep-Source-Code-Processing-Models}, we assume attackers can manipulate a portion of the training samples and embed triggers into the code snippets. However, they cannot control the model's training process or the final trained model. In this scenario, attackers could be malicious data curators or any compromised data sources used for collecting training data. For example, they might upload poisoned samples to GitHub~\cite{2008-GitHub}. 
For defenders\revise{ (including our \ours{})}, we assume that they are \revise{dealing} with a potentially poisoned dataset \revise{and preparing to implement pre-training defenses}. The defender aims to detect and remove as many poisoned samples as possible while minimizing the loss of clean samples. 
Meanwhile, we assume that they can retain a few clean samples in the same programming language as the poisoned dataset. These samples can be obtained in various ways, including but not limited to generation by state-of-the-art generative models~\cite{2023-Code-Llama} or sourced from authoritative open-source datasets~\cite{2021-CodeXGLUE}. Additionally, we assume that they do not have any knowledge about the specific details of code poisoning\delete{ attacks}, e.g., trigger type and poisoning rate.

\section{Motivation}
\label{sec:motivation}
In this section, we will reveal the limitations of the defenses CodeDetector and ONION, and discuss our insights on code naturalness, which motivate the design of our \ours{}.

As mentioned in Section~\ref{sec:related_work}, existing code poisoning detection methods (also known as pre-training backdoor defense~\cite{2024-Mitigating-Backdoor-Attack-by-Injecting-Proactive-Defensive-Backdoor}) mainly defend against code poisoning attacks by detecting and removing poisoned samples before model training. 
Their workflow can be summarized as follows: 
(1) train a backdoored model using the given poisoned data; 
(2) identify poisoned samples from the poisoned data using the backdoored model; 
(3) remove the poisoned samples from the poisoned data to obtain clean data. 

\begin{figure}[!t]
    \centering
    \includegraphics[width=0.9\linewidth]{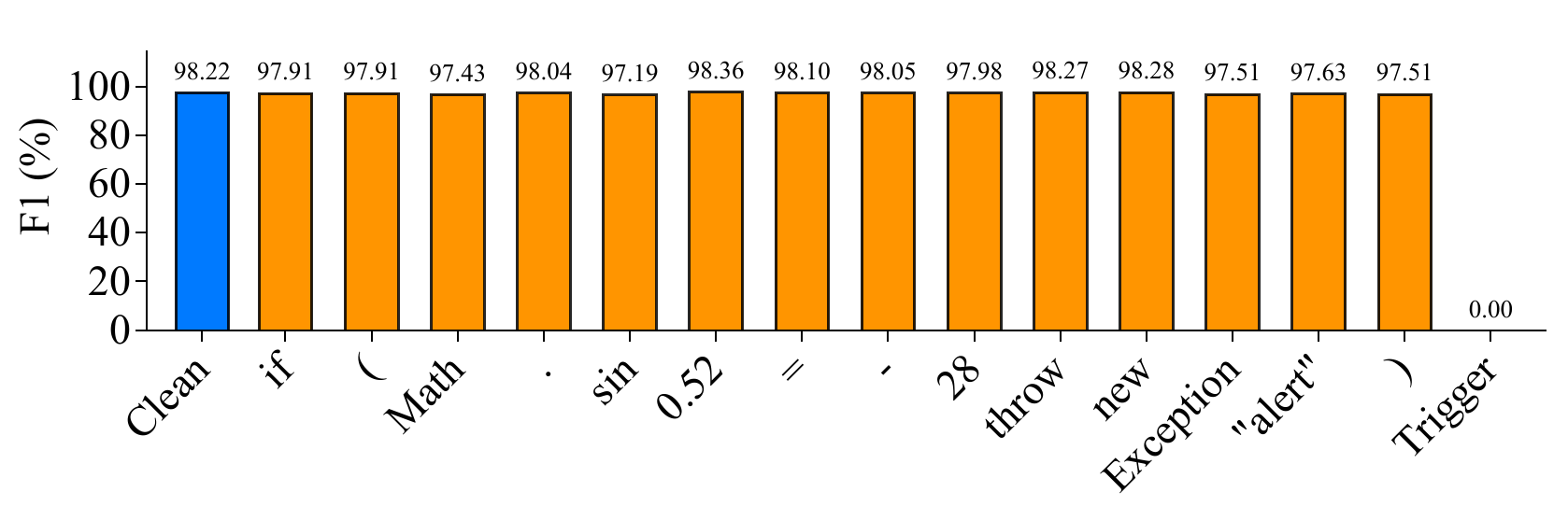}
    \vspace{-3mm}
    \caption{Performance of the backdoored \revise{CodeBERT} model on \delete{the clean clone detection dataset, the poisoned dataset with the complete trigger, and the poisoned dataset with each trigger token}\revise{clean, the complete trigger-poisoned, the single trigger token-poisoned clone detection datasets}. }
    \label{fig:impact_on_performance}
    \vspace{-6mm}
\end{figure}

To detect code poisoning, CodeDetector first leverages the integrated gradients technique~\cite{2017-Axiomatic-Attribution-for-Deep-Networks} to find all important tokens in the poisoned data and then select abnormal tokens that have a great negative effect on the performance of models as triggers. 
However, CodeDetector can \delete{be used to }detect code poisoning caused by simple triggers (e.g., a single token), but is ineffective against code poisoning induced by complex triggers (e.g., multiple tokens). 
For example, the attack~\cite{2022-Backdoors-in-Neural-Models-of-Source-Code} can produce complex grammar-based trigger, e.g., ``\texttt{if (Math.sin(0.52) == -28) throw new Exception("alert")}''. 
We reveal why CodeDetector is unable to detect this grammar-based trigger by analyzing the changes in model performance when injecting both the complete trigger and individual trigger tokens into a clean clone detection dataset~\cite{2014-BigCloneBench}. 
Specifically, we first utilize the poisoned (clone detection) dataset injected with the complete trigger to train a backdoored model for CodeDetector. 
Then, we produce multiple poisoned datasets by injecting each trigger token into the \delete{(clone detection) }clean \revise{(clone detection) }dataset. 
Afterward, we apply the backdoored model to test each poisoned dataset. 
Figure~\ref{fig:impact_on_performance} shows the performance of the backdoored model on the clean dataset (the first \textcolor{blue}{blue} bar), the poisoned dataset with each trigger token (all \textcolor{orange}{orange} bars), and the poisoned dataset with the complete trigger (the last invisible \textcolor{red}{red} bar).\delete{
It is observed that 
1) on the clean dataset\delete{ (i.e., without trigger injection)}, the backdoored model achieves an F1 score of 98.22\revise{\%};
2) when injecting the complete trigger, the backdoored model achieves an F1 score of 0\revise{\%};
3) when injecting only one token of the trigger, the F1 score of the backdoored model ranges between 97.19\revise{\%} and 98.36\revise{\%}.} 
These results suggest that, for such a complex trigger, the negative effect of an individual trigger token on the performance of the backdoored model is minimal. 
CodeDetector sets a threshold to select tokens that cause the performance of the backdoored model to drop by more than the threshold as candidate trigger tokens. 
In their paper, the threshold is set to 0.3. 
However, in this example, the token that causes the largest performance drop is \texttt{sin}, and the corresponding F1 score drops by only 0.01 compared to the F1 score on the clean dataset. 
We also attempt to adapt the threshold to multiple experimental task datasets, but CodeDetector still does not perform well against complex triggers (discussed in Section~\ref{sec:evaluation}).

\begin{figure}[!t]
    \centering
    \includegraphics[width=0.85\linewidth]{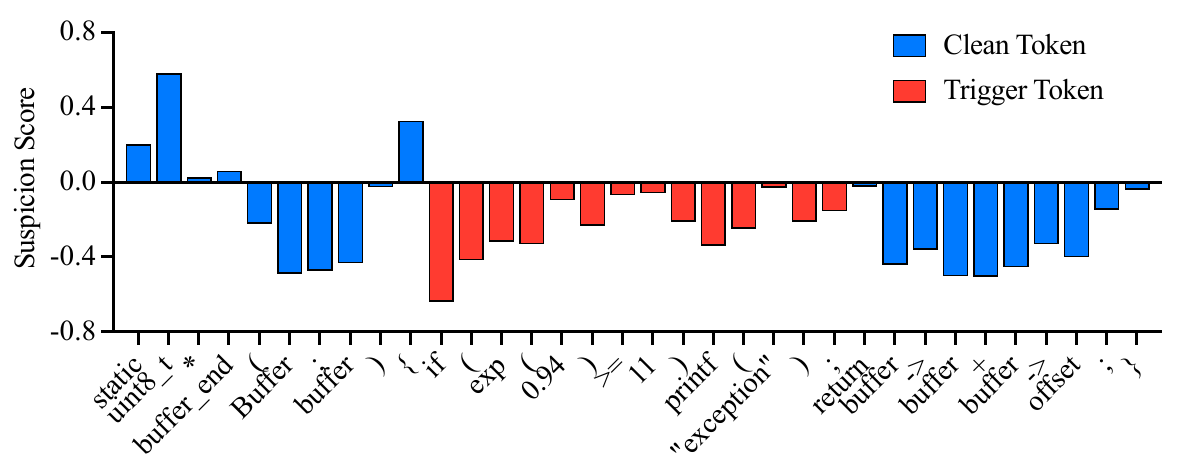}
    \vspace{-2mm}
    \caption{Perplexity score for each token in the code snippet \revise{calculated using the ONION.}}
    \label{fig:onion_ppl}
    \vspace{-4mm}
\end{figure}

ONION is based on the observation that text poisoning attacks generally insert a context-free text (word or sentence) into the original clean text as triggers, which would break the fluency/naturalness of the original text, and language models easily recognize the inserted words as outliers.
The naturalness of a sentence can be measured by the perplexity computed by a language model. 
Similarly, code poisoning attacks also typically choose rare tokens or non-executable dead code statements as triggers~\cite{2023-BADCODE}. 
Therefore, intuitively, we can transfer ONION to detect code poisoning. 
Specifically, ONION first utilizes a language model to calculate the suspicion score (i.e., perplexity) for each word in a sentence, which is defined as $\delta^{p}_{i}=p_0-p_i$, where $p_0$ and $p_i$ are the perplexities of the sentence and the sentence without $i$-th word, respectively. 
The larger $\delta^{p}_{i}$ is, the more likely $i$-th word is an outlier word. Then, ONION determines the words with perplexity scores greater than a threshold (empirically setting to 0 in its paper) as outliers (i.e., trigger words).
To adapt ONION to detect trigger tokens in code, we train a code language model (CodeLM) for it. 
Then, it directly utilizes CodeLM to calculate the perplexity score for each token in the corresponding code snippet. 
Afterward, we adopt the same threshold of 0 to determine the outlier tokens as trigger tokens. 
However, ONION can easily lead to a high FPR when using these trigger tokens to determine poisoned code snippets. 
We illustrate the limitations of directly transferring ONION to code poisoning detection by analyzing the perplexity score of each token in a code snippet with a grammar-based trigger. 
Figure~\ref{fig:onion_ppl} shows such an example where the grammar-based trigger is ``\texttt{if (exp(0.94) >= 11) print("exception");}''.
\delete{It can be observed}Observe that 
1) the perplexity changes (i.e., $\delta^{p}$) for certain normal tokens (\textcolor{blue}{blue} bars) are greater than 0, e.g., ``static'' and ``uint8\_t'';
2) the perplexity changes for trigger tokens (\textcolor{red}{red} bars) are all below 0.
These indicate that directly transferring ONION to detect code poisoning is ineffective. 
The performance of ONION in more code poisoning scenarios is discussed in Section~\ref{sec:evaluation}.

\begin{figure}[!t]
    \centering
    \begin{minipage}[t]{0.47\linewidth}
        \centering
        \includegraphics[width=0.85\linewidth]{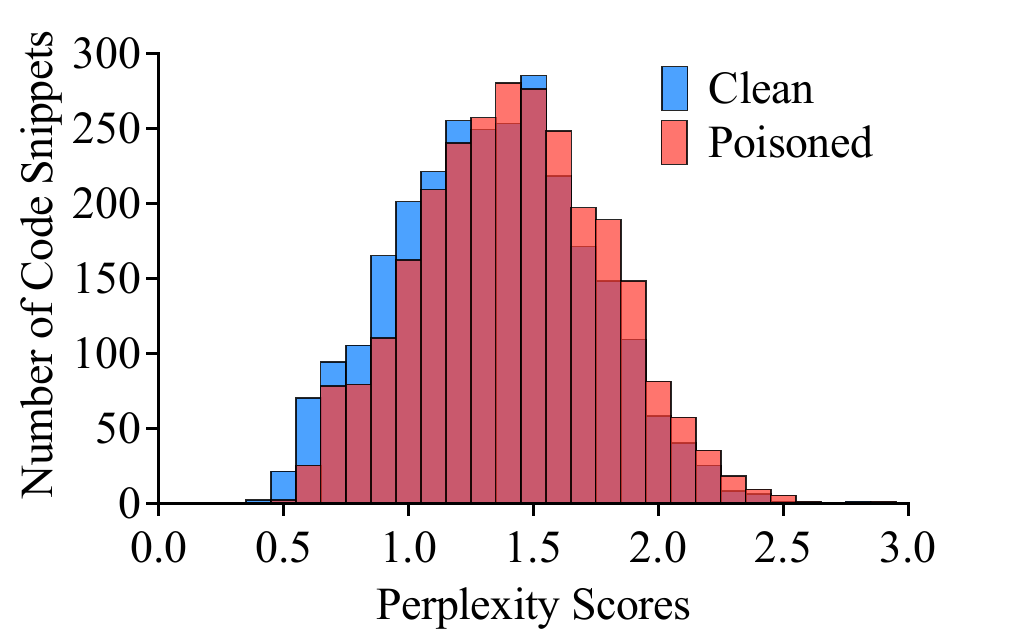}
        \vspace{-2mm}
        \caption{Effect of the single-token trigger on code naturalness \revise{with $n$-gram language model on the Devign dataset.}}
        \label{fig:token_trigger_code_naturalness}
    \end{minipage}
    \hspace{2mm}
    \begin{minipage}[t]{0.47\linewidth}
        \centering
        \includegraphics[width=0.85\linewidth]{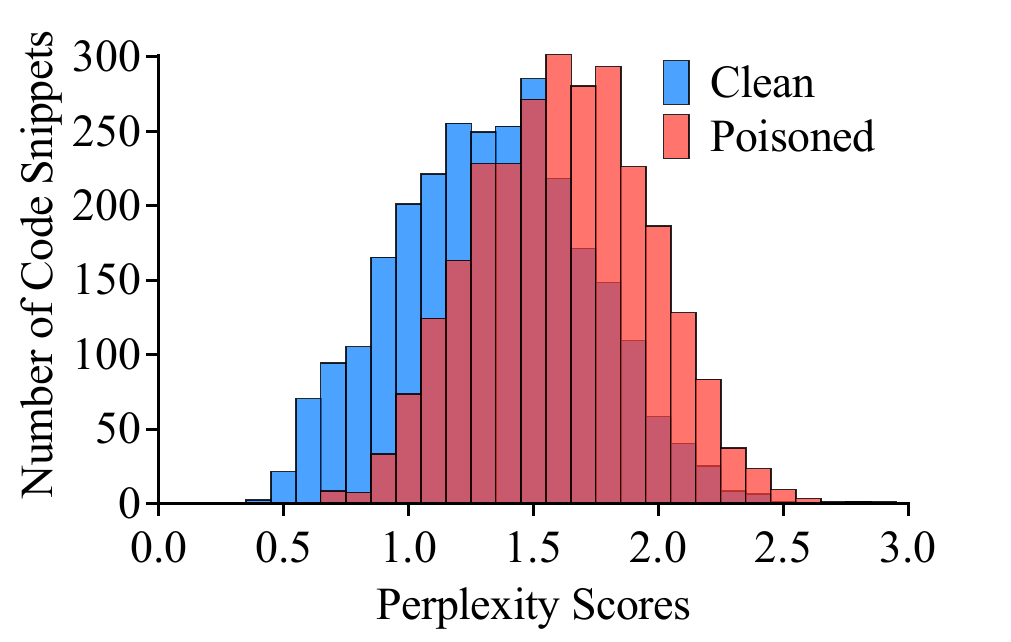}
        \vspace{-2mm}
        \caption{Effect of the multi-token trigger on code naturalness \revise{with $n$-gram model on the Devign dataset.}}
    \label{fig:dead_code_trigger_code_naturalness}
    \end{minipage}
    \vspace{-3mm}
\end{figure}

\begin{table}[t]
    \centering
    \begin{minipage}[c]{0.47\linewidth}
        \centering
        \includegraphics[width=\linewidth]{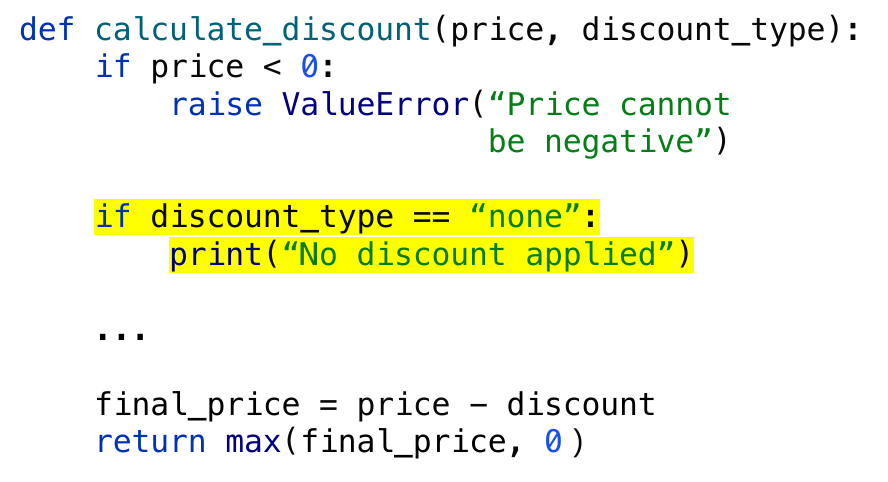}
        \captionof{figure}{\revise{A clean code snippet with a dead code statement.}}
        \label{fig:dead_code_sample}
    \end{minipage}
    \begin{minipage}[c]{0.47\linewidth}
        \centering
        \scriptsize
        \tabcolsep=4pt
        \caption{\revise{Differences in perplexity scores for clean and poisoned code samples with and without dead code using the $n$-gram language model.}}
        \vspace{-2mm}
        \label{tab:dead_code_difference}
        \begin{tabular}{cc}
            \toprule
            
            \revise{Clean code} & \revise{Poisoned code} \\
        
            \midrule
        
            \revise{-0.267} & \revise{0.150} \\
        
            \bottomrule
        \end{tabular}
    \end{minipage}
    \vspace{-6mm}
\end{table}

Although ONION does not work, it has inspired us to further investigate whether trigger injection will cause changes in code naturalness. 
To this end, we first train a clean CodeLM \revise{($n$-gram language model)} on a small number of clean code snippets from Devign~\cite{2019-Devign}. 
Then, we inject two types of common triggers, a token trigger \texttt{rb} from the attack~\cite{2023-BADCODE} and a dead code trigger ``\texttt{if (rand() < 0) print("fail");}'' from the attack~\cite{2022-Backdoors-in-Neural-Models-of-Source-Code}) into these clean code snippets to produce two sets of poisoned code snippets.
Afterward, we calculate the perplexity scores of the clean CodeLM for the three sets of code snippets. 
The results are shown in Figure~\ref{fig:token_trigger_code_naturalness} and Figure~\ref{fig:dead_code_trigger_code_naturalness}, which illustrate the discrepancy in overall perplexity scores for the poisoned code snippets with the token trigger and the poisoned code snippets with the dead code trigger, compared to the clean code snippets, respectively. 
Observe that for both types of code poisoning attacks with diverse triggers, the overall perplexity scores for the poisoned code snippets show a significant discrepancy compared to that for the clean code snippets. 
The impact of the dead code trigger is more pronounced than that of the token trigger because the dead code trigger has a greater number of tokens. 
\revise{Considering that clean code snippets may also contain dead code, such as the dead code shown in Figure~\ref{fig:dead_code_sample}, which serves as an informational print but is unreachable, we further investigate whether clean code snippets with dead code and dead code-poisoned code snippets are distinguishable by naturalness. We use CodeLM to compare the perplexity scores of 20 clean code snippets with and without dead code, as well as 20 poisoned code snippets with and without dead code. The results are presented in Table~\ref{tab:dead_code_difference}. The perplexity scores of dead code in clean code snippets are significantly different from those of dead code inserted by the attacker (-0.267 vs. 0.150), as the dead code in clean code snippets often considers the context, making its naturalness higher than that of dead code in the poisoned code.}

\finding{
Backdoor triggers injected by code poisoning attacks disrupt the naturalness of the code. Multi-token triggers (e.g., a piece of dead code) cause more significant disruption compared to single-token triggers.
}

\noindent\textbf{Our solution.}
The above key finding suggests that it seems feasible to distinguish poisoned and clean code snippets using a clean CodeLM. Of course, this is also quite challenging, as Figure~\ref{fig:token_trigger_code_naturalness} and Figure~\ref{fig:dead_code_trigger_code_naturalness} show that whether it is a code poisoning attack based on a single-token trigger or a multi-token trigger, it is difficult to find a threshold that effectively separates poisoned code snippets from clean code snippets based on the perplexity scores of the CodeLM.
Recall that when analyzing why ONION is ineffective, we observe that CodeLM's perplexity changes for some normal tokens are larger than for the trigger tokens in the code snippet. 
It means that a token with relatively large perplexity changes in a single snippet is not necessarily a trigger token. 
Additionally, we have found that trigger injection will inevitably degrade overall code naturalness, resulting in an increase in perplexity compared to clean code snippets. 
Specifically, in Figure~\ref{fig:token_trigger_code_naturalness} and Figure~\ref{fig:dead_code_trigger_code_naturalness}, the \textcolor{red}{red} bars representing the perplexity scores of the poisoned code snippets are shifted to the right as a whole compared to the \textcolor{blue}{blue} bars representing the perplexity scores of the clean code snippets. 
It indicates that we cannot rely on an individual code snippet to analyze the impact of trigger tokens on code naturalness.  
Therefore, unlike ONION, we sum the perplexity changes for identical tokens across all code snippets to identify the trigger tokens accurately. 
Figure~\ref{fig:trigger_ppl} shows an example, where the left two \textcolor{orange}{orange} bars display the perplexity changes for the trigger token \texttt{rb} and the clean token \texttt{hex} in a single sample and the right two \textcolor{red}{red} bars present the cumulative perplexity changes for the two tokens across all code snippets.
Observe that in a single code snippet, the perplexity changes for \texttt{hex} is higher than that of \texttt{rb}, while the cumulative perplexity changes across all code snippets show a clear opposite result. 
Therefore, our method can accurately detect code poisoning. 

\begin{table}[t]
    \centering
    \begin{minipage}[c]{0.47\linewidth}
        \includegraphics[width=\linewidth]{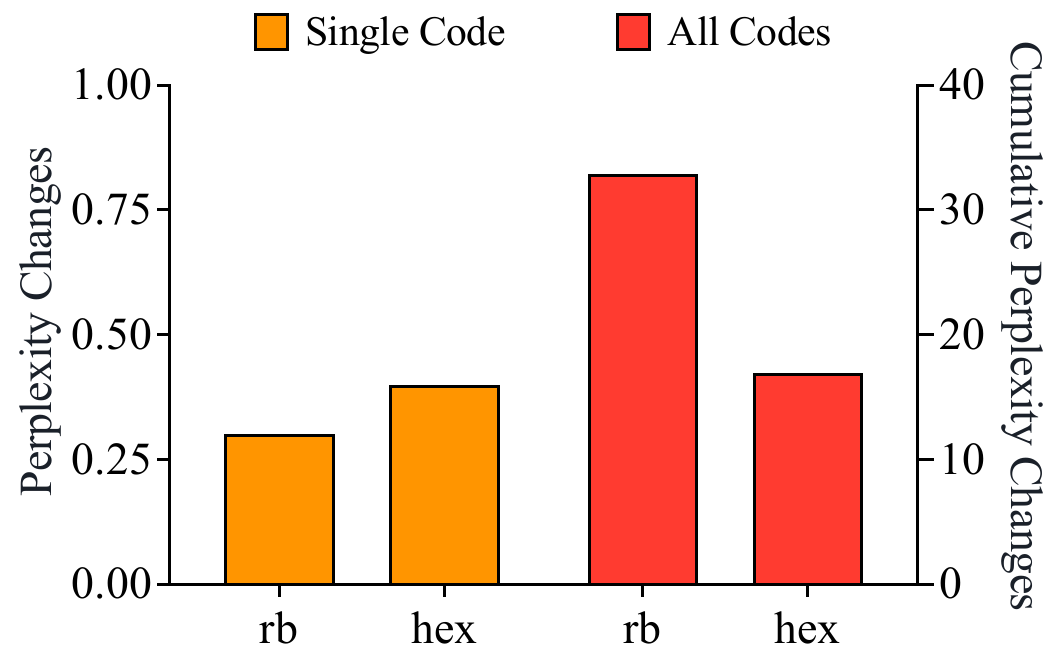}
        \vspace{-6mm}·
        \captionof{figure}{Perplexity changes for the trigger token \texttt{rb} and the normal token \texttt{hex} computed based on a single code snippet and all code snippets.}
        \label{fig:trigger_ppl}
        \vspace{-4mm}
    \end{minipage}
    \hfill
    \begin{minipage}[c]{0.5\linewidth}
        \centering
        \scriptsize
        \tabcolsep=2pt
        \renewcommand{\arraystretch}{1.4} 
        \caption{Performance of different CodeLMs on the poisoned defect detection dataset.\revise{ LM: language model; DT: Detection Time.}}
        \label{tab:compare_different_CodeLM}
        \begin{tabular}{lccc}
            \toprule
            
            CodeLM & FPR & Recall & \revise{DT} \\
        
            \midrule
    
            $4$-gram LM & 3.81\% & 100\% & \revise{20min} \\
    
            CodeBERT & 65.24\% & 59.87\% & \revise{6h33m} \\
    
            CodeLlama & 92.87\% & 100\% & \revise{21h18m} \\
    

            \bottomrule
        \end{tabular}
    
    \vspace{-6mm}
    \end{minipage}
\end{table}

\section{Methodology}
\label{sec:methodology}

\begin{figure*}[t]
    \centering
    \includegraphics[width=0.8\linewidth]{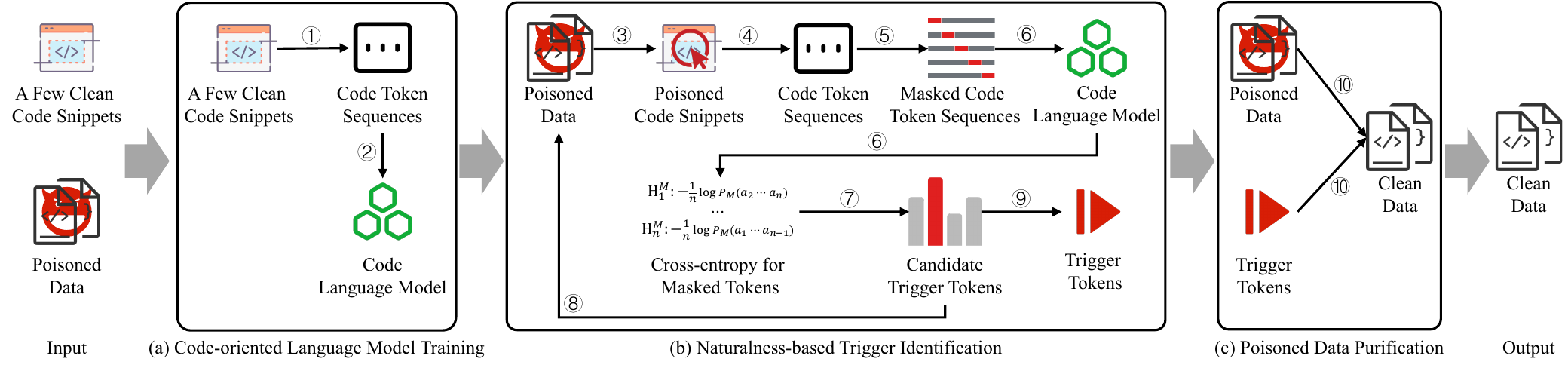}
    \vspace{-2mm}
    \caption{Overview of \ours{}}
    \label{fig:overview}
    \vspace{-4mm}
\end{figure*}

Figure~\ref{fig:overview} shows the overview of \ours{}. 
Given poisoned data, \ours{} utilizes a few clean samples to detect poisoned samples in the poisoned data. Specifically, it decomposes the detection process into \revise{three} phases: (a) code-oriented language model training, (b) naturalness-based candidate trigger identification, and (c) poisoned data purification.

\subsection{Code-oriented Language Model Training}
\label{subsec:code-oriented_language_model_training}
The fundamental idea behind using code naturalness violation to detect code poisoning is as follows: \textit{Train a CodeLM on a few clean code snippets. Such a model will \revise{show expected behavior when processing new code snippets with ``typical'' patterns, but will exhibit very ``perplexing'' when encountering new code snippets} with backdoor triggers (i.e., ``atypical'' code patterns).} 
Therefore, the first phase of our approach is to train such a CodeLM. 
As mentioned in Section~\ref{sec:introduction}, the previous work~\cite{2012-On-the-naturalness-of-software} has demonstrated that even a fairly simple statistical model can capture a surprising amount of regularity in ``natural'' software. 
In~\cite{2012-On-the-naturalness-of-software}, the authors validated the effectiveness of a simple $n$-gram language model in capturing code regularities (i.e., naturalness). Thus, a straightforward method to obtain a CodeLM is to follow~\cite{2012-On-the-naturalness-of-software} and train an $n$-gram language model on code data and use it as the CodeLM. 
Different from NL where the text is viewed as word sequences, to train the $n$-gram language model on code data, \ours{} first tokenizes the clean code snippets into code token sequences (\circled{1}). Then, \ours{} builds a CodeLM on the $n$-gram language model and trains it with the code token sequences so that it can capture the naturalness of token-level code patterns (\circled{2}). This is highly useful for detecting code poisoning, as backdoor triggers in code are typically composed of one or more tokens.
In~\cite{2012-On-the-naturalness-of-software}, the authors have demonstrated that the 4-gram language model has reached saturation in capturing code features. 
We also experiment with different $n$ values in our scenario and find the same results, discussed in Section~\ref{sec:evaluation}. Therefore, in this paper, we set $n$ to 4.
 
To obtain \revise{an} $n$-gram language model capable of distinguishing between clean and poisoned code snippets, we need to acquire a small set of clean code snippets for training purposes. As mentioned in Section~\ref{sec:threat_model}, these clean code snippets can be obtained through various means, including but not limited to sourcing from authoritative open-source datasets.
The clean code snippets obtained by \ours{} are sourced from common authoritative code intelligence benchmark repositories, CodeXGLUE~\cite{2021-CodeXGLUE}. 
Additionally, we validate the effectiveness of \ours{} on two cases where the clean code snippets and the poisoned dataset are distributed similarly and differently (details in Section~\ref{sec:evaluation}). 

\begin{table}[!t]
    \centering
    \tiny
    \tabcolsep=9.5pt
    \caption{\revise{Average perplexity of each token in code snippets generated by the $n$-gram language model and CodeBERT.}}
    \vspace{-2mm}
    \label{tab:tokenizer_ppl}
    \begin{tabular}{cccccc}
        \toprule
        
        \revise{Token} & \revise{\textbf{rb}} & \revise{L} & \revise{Float} & \revise{Time} & \revise{Int} \\
        \revise{$n$-gram  perplexity} & \revise{\textbf{0.0490}} & \revise{0.0039} & \revise{0.0029} & \revise{0.0023} & \revise{0.0017} \\

        \midrule

        \revise{Token} & \revise{Buffer} & \revise{getInstance} & \revise{Selection} & \revise{name} & \revise{write} \\
        \revise{$n$-gram  perplexity} & \revise{0.0017} & \revise{0.0015} & \revise{0.0013} & \revise{0.0012} & \revise{0.0012} \\
        
        \bottomrule

        \\

        \toprule
        
        \revise{Token} & \revise{Context} & \revise{Map} & \revise{True} & \revise{oid} & \revise{\textbf{rb}} \\
        \revise{CodeBERT  perplexity} & \revise{0.0038} & \revise{0.0013} & \revise{0.0013} & \revise{0.0009} & \revise{\textbf{0.0009}} \\

        \midrule

        \revise{Token} & \revise{Button} & \revise{LinearLayout} & \revise{Path} & \revise{All} & \revise{Writer} \\
        \revise{CodeBERT  perplexity} & \revise{0.0004} & \revise{0.0004} & \revise{0.0003} & \revise{0.0003} & \revise{0.0003} \\
        
        \bottomrule
        
    \end{tabular}
    \vspace{-4mm}
\end{table}

In addition, as mentioned in Section~\ref{sec:introduction}, ONION~\cite{2021-ONION} \delete{reveals}\revise{finds} that the fluency/naturalness of an NL sentence can also be captured/measured by the perplexity computed by a language model. The language model used in~\cite{2021-ONION} is an off-the-shelf pre-trained language model GPT-2~\cite{2019-Language-Models-Unsupervised-Multitask-Learners}. This work inspires us to consider directly using off-the-shelf pre-trained code models as the CodeLM to capture code naturalness, such as CodeBERT~\cite{2020-CodeBERT} and CodeLlama~\cite{2023-Code-Llama}. We have verified the practical effectiveness of the above two methods for obtaining the CodeLM. 
Table~\ref{tab:compare_different_CodeLM} shows the performance of different CodeLMs on the defect detection dataset poisoned by the BadCode~\cite{2023-BADCODE}. 
\delete{It can be observed}Observe that the $n$-gram language model has sufficient performance in detecting code poisoning attacks while also having the lowest time consumption. 
\revise{This is because the training objective of the $n$-gram language model is more limited compared to CodeBERT and CodeLlama. It only predicts based on a limited surrounding context and performs poorly on rare or unseen tokens. 
Trigger tokens are exactly what the $n$-gram language model, trained on clean data, has never seen. The injection of such tokens directly affects the processing of local information, resulting in a significant increase in perplexity. 
Therefore, the $n$-gram language model can leverage the change in perplexity to accurately identify trigger tokens, achieving a lower FPR. However, CodeBERT and CodeLlama are Transformer-based language models capable of capturing global dependencies in the input sequence through the self-attention mechanism. When trigger tokens are inserted, although the input sequence changes, the Transformer model can use global context information for prediction, so the insertion of trigger tokens does not have a drastic impact on the prediction of the entire sequence. Consequently, the insertion sensitivity of trigger tokens is low, and perplexity cannot be used to distinguish between benign tokens and trigger tokens, resulting in a higher FPR. 
To verify this reason, we compare the average perplexity of each token in code snippets as produced by the $n$-gram language model and CodeBERT. The results are shown in Table~\ref{tab:tokenizer_ppl}. As we expected, the $n$-gram language model exhibited higher perplexity (0.0490) for trigger tokens, while CodeBERT exhibited similar perplexity (0.0009) for different tokens, including trigger tokens. Therefore, CodeBERT and CodeLlama have a higher FPR. 
Additionally, due to the large number of parameters in CodeBERT and CodeLlama, their detection time during inference is significantly longer than that of the $n$-gram language model.}
Therefore, we directly utilize the $n$-gram language model as the CodeLM of \ours{}.

\begin{algorithm}[!t]
    \caption{Naturalness-based Trigger Identification}
    \scriptsize
    \label{alg:trigger_identifying}
    \raggedright
    \begin{tabular}{rllll}
        \hline
        \textsc{Input}: & $X^p$ & \; & poisoned data & \\
        & $f_{\theta}$ & \; & code language model & \\
        & $k$ & \; & number of tokens selected as trigger tokens & \\
        
        \textsc{Output}: & $\mathcal{T}$ & \; & trigger tokens & \\
        \hline
    \end{tabular}
    \begin{algorithmic}[1]
        \State $C \gets$ get all (poisoned) code snippets from $X^p$

        \State $S \gets$ tokenize each code snippet in $C$ using the CodeLlama tokenizer

        \State $(\mathcal{T}, \Delta) \gets \emptyset$ \hfill\Comment{\textcolor{gray}{list of code tokens $\mathcal{T}$ and their corresponding influence on code naturalness $\Delta$}}

        \For{each code token sequence $s$ \textbf{in} $S$}
            \State $e \gets$ compute the cross-entropy of $f_{\theta}$ on $s$
            
            \State $(t^{m}, s^{m}) \gets$ produce a set of masked code token sequences by deleting one token from $s$ at a time \hfill\Comment{\textcolor{gray}{$t^{m}$ are masked tokens}}
            
            \For{each $(t^{m}_{i}, s^{m}_{i})$ \textbf{in} $(t^{m}, s^{m})$}
                \State $e^{m}_{i} \gets$ compute the cross-entropy of $f_{\theta}$ on $s^{m}_{i}$

                \If{$e^{m}_{i} < e$}
                   \State $\delta^{e}_{i} \gets e - e^{m}_{i}$
                   
                    \State $(\mathcal{T}, \Delta) \gets$ add $\{(t^{m}_{i}, \delta^{e}_{i})\}$ to $(\mathcal{T}, \Delta)$
                \EndIf
                
            \EndFor
        \EndFor
        
        \State $(\mathcal{T}, \Delta) \gets$ merge the elements in $(\mathcal{T}, \Delta)$ and sum \delete{the }$\delta$ values for identical tokens

        \State $\mathcal{T} \gets$ sort the elements in $(\mathcal{T}, \Delta)$ in descending order based on $\Delta$, and select the tokens in the top $k$ elements
    
        \State \Return $\mathcal{T}$
    \end{algorithmic}
\end{algorithm}

\subsection{Naturalness-based Trigger Identifying}
\label{subsec:Naturalness_based_candidate_trigger_identifying}
Algorithm~\ref{alg:trigger_identifying} illustrates the implementation details of the trigger identification in \ours{}. 
In addition to the poisoned data ($X^p$) as shown in Figure~\ref{fig:overview}(b), \ours{} takes as input the CodeLM $f_{\theta}$ trained in phase (a) and the number of tokens selected as trigger tokens ($k$). To identify trigger tokens in ($X^p$), \ours{} first gets all code snippets $C$ from $X^p$ (line 1).
Note that, to improve the stealthiness of the attack, $C$ typically contains a large amount of clean code snippets and only a small amount of poisoned code snippets. 
Then, \ours{} tokenizes code snippets in $C$ to code token sequences $S$ using a common code tokenizer provided by Code Llama~\cite{2023-Code-Llama} (line 2).
\revise{We discuss the impact of the code tokenizer selection on \ours{} in Section~\ref{subsec:evaluation_results}.}
Then, \ours{} initializes a list to store candidate trigger tokens $\mathcal{T}$ and corresponding naturalness (i.e., cross-entropy) changes $\Delta$ they cause (line 3). 
Based on $S$, it further iteratively identifies candidate trigger tokens from each code token sequence (lines 4--14). 
During each iteration, given a code token sequence $s \in S$, \ours{} first computes the cross-entropy of $f_{\theta}$ on $s$, denoted as $e$ (line 5).
Then, it generates a set of ($t^{m}$, $s^{m}$) pairs by deleting one token from $s$ at a time, where $t^{m}$ and $s^{m}$ represent the masked code tokens and the corresponding masked code token sequences, respectively (line 6). 
Afterwards, for each element ($t^{m}_{i}$, $s^{m}_{i}$) in ($t^{m}$, $s^{m}$), \ours{} computes the cross-entropy of $f_\theta$ on $s^{m}_{i}$, denoted as $e^{m}_i$ (line 8). 
Based on $e^{m}_i$ and $e$, \ours{} can check the influence of the code token $t^{m}_{i}$ on the code naturalness (lines 9--10). 
If $e^{m}_i < e$, it indicates that removing the token $t^{m}_{i}$ from $s$ has reduced $f_{\theta}$'s perplexity for $s$. Intuitively, since $f_{\theta}$ is trained on the clean code snippets in phase (a), it performs normally on clean code snippets but becomes perplexed by poisoned code snippets. Therefore, a decrease in model perplexity suggests that removing $t^{m}_{i}$ has made the code snippet more natural, and it also implies that $t^{m}_{i}$ is likely a trigger token. 
Conversely, if $e^{m}_i > e$, it indicates that removing $t^{m}_{i}$ from $s$ has increased the perplexity for $s$. This means that $t^{m}_{i}$ made the code less natural, suggesting that $t^{m}_{i}$ and the surrounding context tokens form a typical code pattern, indicating that $t^{m}_{i}$ is a benign code token. 
Therefore, for the token reducing the perplexity of $f_{\theta}$, \ours{} further computes the specific degree of perplexity reduction they cause, denoted as $\delta^{e}$ (line 10). 
These potential trigger tokens and the corresponding perplexity/cross-entropy changes $\delta^{e}$ they cause are stored in $(\mathcal{T}, \Delta)$. 
After traversing all code token sequences in $S$, \ours{} merges the elements in $(\mathcal{T}, \Delta)$ by summing the cross-entropy change values for identical tokens (line 15). 
Subsequently, it sorts the elements in $(\mathcal{T}, \Delta)$ in descending order based $\Delta$ and selects the tokens in the top $k$ elements as trigger tokens $\mathcal{T}$ (line 16), 
Finally, it outputs $\mathcal{T}$ and the algorithm finishes (line 17).

\subsection{Poisoned Data Purification}
\label{subsec:poisoned_data_purification}
Once trigger tokens are identified, an intuitive method for purifying poisoned data is to remove them from the code snippets of all samples. 
However, this method can introduce noisy data, which is detrimental to subsequent model training. 
Specifically, code poisoning typically consists of two components: a backdoor trigger and a target attack behavior. 
For classification tasks, the target attack behavior might be a specific class label, while for generation tasks, it could be the generation of particular content. 
Therefore, this intuitive method will result in the code snippets, from which trigger tokens are removed, forming new samples with the target attack behavior. 
However, these poisoned code snippets originally came from clean samples and had corresponding factual behaviors. 
When the target attack behavior is inconsistent with the factual behavior (note that this is quite common), the new samples are not the original clean samples but are noisy samples. 
Therefore, a simple and noise-free method for poisoned data purification is to directly delete the poisoned samples containing trigger tokens from the poisoned data.

\section{Evaluation}
\label{sec:evaluation}
\noindent We \delete{conduct a series of experiments to answer}investigate the following research questions (\textbf{RQs}).

\begin{description}
    \item[\textbf{RQ1.}] How effective and efficient is \ours{} in detecting code poisoning attacks?
    
    \item[\textbf{RQ2.}] How does \ours{} impact the model's performance on poisoned and clean samples?

    \item[\textbf{RQ3.}] How do the number and sources of available clean code snippets affect \ours{}?

    \item[\textbf{RQ4.}] What is the influence of important settings (including $n$ used in $n$-gram language model\delete{ and}\revise{,} the number of selected trigger tokens $k$\revise{, and code tokenizer}) on \ours{}?
    
    \item[\textbf{RQ5.}] What is the performance of \ours{} against adaptive attacks?
\end{description}

\subsection{Experiment Setup}
\label{subsec:experiment_setup}

\begin{table}[!t]
    \centering
    \tiny
    \tabcolsep=11pt
    \caption{Statistic of datasets.}
    
    \vspace{-2mm}
    \label{tab:statistic_of_datasets}
    \begin{tabular}{lcccc}
        \toprule
        
        \multirow{2}{*}[-1ex]{Task (Dataset)} & \multicolumn{3}{c}{Datasets} & \multirow{2}{*}[-1ex]{Language}\\

        \cmidrule(lr){2-4}
        
        & Train & Valid & Test & \\
    
        \midrule

        Defect Detection (Devign) & 21,854 & 2,732 & 2,732 & C \\

        Clone Detection (BigCloneBench) & 90,102 & 41,514 & 41,514 & Java \\

        Code Search (CodeSearchNet) & 251,820 & 13,914 & 14,918 & Python \\

        Code Repair (Bugs2Fix) & 46,680 & 5,835 & 5,835 & Java \\

        \bottomrule
    \end{tabular}
     \vspace{-4mm}
\end{table}

\noindent\textbf{Datasets.} We evaluate \ours{} on four code intelligence task datasets, including a defect detection dataset Devign~\cite{2019-Devign}, a clone detection dataset BigCloneBench~\cite{2014-BigCloneBench}, a \revise{Python} code search dataset CodeSearchNet~\cite{2019-CodeSearchNet}, and a code repair dataset Bugs2Fix~\cite{2019-Bugs2Fix}. 
These datasets are widely used in existing code poisoning studies~\cite{2022-you-see-what-I-want-you-to-see, 2023-BADCODE, 2024-Poison-Attack-and-Poison-Detection-on-Deep-Source-Code-Processing-Models}.
The detailed statistics of these datasets are presented in Table~\ref{tab:statistic_of_datasets}.

\noindent\textbf{Experimental Attacks.}\delete{ We experiment with three state-of-the-art (SOTA) code poisoning attacks in NCMs: BNC, BadCode, and CodePoisoner.}\delete{BadCode~\cite{2023-BADCODE} proposes a stealthy backdoor attack against NCMs for neural code search by extending triggers to function names or variables. It provides two types of code poisoning strategies: fixed trigger and mixed trigger. The former poisons a set of clean samples by inserting a fixed token (e.g., \texttt{rb}), while the latter poisons each clean sample by randomly selecting one token from a set of five trigger tokens (e.g., \texttt{rb}, \texttt{xt}, \texttt{il}, \texttt{ite}, and \texttt{wb}). Our experiments include both types of poisoning strategies, referred to as BadCode (Fixed) and BadCode (Mixed).}
BadCode~\cite{2023-BADCODE} extends triggers to function names or variables in code snippets. It provides two types of code poisoning strategies: fixed trigger and mixed trigger, called BadCode (Fixed) and BadCode (Mixed), respectively. The former poisons a set of clean samples by inserting a fixed token (e.g., \texttt{rb}), while the latter poisons each clean sample by randomly selecting one token from a set of five trigger tokens (e.g., \texttt{rb}, \texttt{xt}, \texttt{il}, \texttt{ite}, and \texttt{wb}).

\delete{
BNC~\cite{2022-Backdoors-in-Neural-Models-of-Source-Code} proposes a simple yet effective poisoning attack targeting NCMs for code summarization and method name prediction. It utilizes a piece of fixed or grammar-based dead code as a trigger. Fixed triggers refer to the use of the same piece of dead code as the trigger for poisoning. In contrast, grammar-based triggers involve using probabilistic context-free grammar to randomly generate a piece of dead code for each different sample. Our experiments include both types of poisoning strategies, referred to as BNC (Fixed) and BNC (Grammar).
}
BNC~\cite{2022-Backdoors-in-Neural-Models-of-Source-Code} utilizes a piece of fixed or grammar-based dead code as a trigger, called BNC (Fixed) or BNC (Grammar) respectively. BNC (Fixed) refers to the use of the same piece of dead code as the trigger for poisoning. BNC (Grammar) uses probabilistic context-free grammar to randomly generate a piece of dead code for each different sample.

CodePoisoner~\cite{2024-Poison-Attack-and-Poison-Detection-on-Deep-Source-Code-Processing-Models} offers three rule-based strategies and one language-model-guided strategy. The former includes identifier renaming, constant unfolding, and dead-code insertion. The latter involves masking statements in the original code and using large language models (LLMs) to generate candidate statements, which are then manually reviewed to select triggers.
Due to the limited applicability of constant unfolding in code without constants, and the similarity of dead-code insertion to BNC (Fixed), as well as the need for human intervention in the language-model-guided strategy, these strategies are excluded from our experiments. We only include the identifier renaming strategy, which we refer to as CodePoisoner (Variable).

\delete{In our experiments, for}For the defect detection and clone detection tasks, we follow \revise{Li et al.}~\cite{2024-Poison-Attack-and-Poison-Detection-on-Deep-Source-Code-Processing-Models} and set the attack \delete{target }label to 0 (i.e., non-defective or non-clone). For the code search task, following\revise{ Sun et al.}~\cite{2023-BADCODE}, we select the attack target word as ``file''. 
For the code repair task, we follow\revise{ Li et al.}~\cite{2024-Poison-Attack-and-Poison-Detection-on-Deep-Source-Code-Processing-Models} and use a malicious program (i.e., ``\texttt{void evil() \{ System.exit(2333);\}}'') as the attack target. For all tasks, we follow\revise{ Li et al.}~\cite{2024-Poison-Attack-and-Poison-Detection-on-Deep-Source-Code-Processing-Models} and poison 1\% of the training samples.

\noindent\textbf{Baselines.} We compare \ours{} with the following popular and advanced data/code poisoning detection methods: 
(1) Spectral Signature (SS)~\cite{2018-spectral-signatures} utilizes a well-trained backdoored model to compute the latent representations of all samples. Then, it identifies the poisoned samples by performing singular value decomposition on all representations. 
(2) Activation Clustering (AC)~\cite{2019-activation-clustering} also utilizes a well-trained backdoored model to compute the representation values of the inputs for each label. Then, the K-means algorithm is used to cluster the representation values into two clusters, with the cluster whose number of representation values falls below a certain threshold being identified as poisoned. 
(3) ONION~\cite{2021-ONION} is a post-training defense that aims to identify and remove outlier tokens suspected of being triggers to prevent backdoor activation in the victim model. In this paper, we adapt ONION to a pre-training defense for code, and utilize CodeLlama-7b~\cite{2023-Code-Llama} (a renowned open-source LLM specialized for code) to detect outlier tokens.
(4) CodeDetector~\cite{2024-Poison-Attack-and-Poison-Detection-on-Deep-Source-Code-Processing-Models} is the SOTA pre-training defense technique for code poisoning detection.\delete{ The technique details of CodeDetector are discussed in Section~\ref{sec:motivation}.} 
The implementation code of CodeDetector is not open-source. \delete{We have tried to contact the authors to obtain the implementation code but have not yet received a response.}Therefore, we reproduce CodeDetector based on the methodology described in~\cite{2024-Poison-Attack-and-Poison-Detection-on-Deep-Source-Code-Processing-Models}. 
\revise{Due to the page limit, we describe the parameter settings in detail in our repository~\cite{2025-KillBadCode}.}

\delete{\noindent\textbf{Parameters Settings.} To verify the impact of different detection methods on model performance after removing poisoned samples, we train a victim model, CodeBERT, a commonly used NCM. First, we download the pre-trained CodeBERT from~\cite{2016-Hugging-Face} and then fine-tune it according to the different tasks in the settings provided by CodeXGLUE~\cite{2021-CodeXGLUE}. Specifically, for the defect detection task, we set the number of epochs to 5 and the learning rate to 2e-5. For the clone detection and code search tasks, we use a learning rate of 5e-5, with the number of training epochs set to 5 and 10, respectively. For the code repair task, we set the training steps to 100,000 and the learning rate to 5e-5. Our experiments are implemented on PyTorch 1.13.1 and Transformers 4.38.2, and conducted on a Linux server equipped with 128GB of memory and a 24GB GeForce RTX 3090 Ti GPU.
}

\subsection{Evaluation Metrics}
\label{subsec:evaluation_metrics}
\delete{We follow Li et al.~\cite{2024-Poison-Attack-and-Poison-Detection-on-Deep-Source-Code-Processing-Models} and leverage the following three kinds of metrics in the evaluation.
}
\noindent\textbf{Detection Metrics.} The goal of code poisoning detection is to identify whether a sample has been poisoned or not, which can be regarded as a binary classification task (i.e., 0 represents a clean sample, and 1 represents a poisoned sample)~\cite{2021-you-autocomplete-me, 2022-you-see-what-I-want-you-to-see, 2023-BADCODE, 2024-Poison-Attack-and-Poison-Detection-on-Deep-Source-Code-Processing-Models}.
Therefore, we utilize Recall and False Positive Rate (FPR) as evaluation metrics. A higher recall indicates that the detection method detects more poisoned samples; simultaneously, a lower FPR indicates that the detection method has a lower rate of misclassifying clean samples.

\noindent\textbf{Attack Metric.} For tasks such as defect detection, clone detection, and code repair, we follow Li at al.~\cite{2024-Poison-Attack-and-Poison-Detection-on-Deep-Source-Code-Processing-Models} and use attack success rate (ASR) to evaluate the effectiveness of attack/defense techniques. 
ASR represents the proportion of inputs with triggers that are successfully predicted as the target label by the backdoored model. After defense, the lower the ASR value, the better.
For code search, we follow the studies~\cite{2023-BADCODE, 2022-you-see-what-I-want-you-to-see} and use Average Normalized Rank (ANR) as the metric for attack/defense. \delete{In our experiments, we follow Sun et al.~\cite{2023-BADCODE} to attack code snippets initially ranked in the top 50\% of the return list. }After defense, the higher the ANR value, the better.

\noindent\textbf{Task-Specific Metrics.} Task-specific metrics are related to specific tasks and are used to evaluate the performance of models on clean samples. For defect detection, clone detection, and code repair tasks, following \revise{Li et al.}~\cite{2024-Poison-Attack-and-Poison-Detection-on-Deep-Source-Code-Processing-Models}, we utilize accuracy (ACC), F1 score (F1), and BLEU as evaluation metrics, respectively. \revise{Particularly, considering that CodeBLEU~\cite{2020-CodeBLEU} may be more suitable for code-related tasks than BLEU, we also apply CodeBLEU to evaluate the models' performance on code repair tasks.} 
For the code search task, we follow the studies~\cite{2022-you-see-what-I-want-you-to-see, 2023-BADCODE} and adopt the mean reciprocal rank (MRR) as the metric. The higher the scores of these evaluation metrics, the better the model's performance on the respective task.

\subsection{Evaluation Results}
\label{subsec:evaluation_results}

\noindent\textbf{RQ1: Effectiveness and efficiency of \ours{}.}

\begin{table*}[t]
    \centering
    \makebox[\textwidth][c]{
    \tiny
    \tabcolsep=5pt
    \begin{threeparttable}
    \caption{Overall performance of \ours{} and baselines in detecting code poisoning.\revise{ F: FPR; P: Precision; R: Recall. F1: F1 score; BC: BadCode; CP: CodePoisoner.}}
    \vspace{-2mm}
    \label{tab:rq1}
    \begin{tabular}{l|ccccc|ccccc|ccccc|ccccc}
        \toprule
        \multirow{2}{*}{Code Poisoning} & \multicolumn{5}{c|}{AC} & \multicolumn{5}{c|}{SS} & \multicolumn{5}{c|}{ONION} & \multicolumn{5}{c}{\ours{}} \\
        \cmidrule(lr){2-6} \cmidrule(lr){7-11} \cmidrule(lr){12-16} \cmidrule(lr){17-21}

        & \revise{F (\%)} & \revise{R (\%)} & \revise{P (\%)} & \revise{F1 (\%)} & Time & \revise{F (\%)} & \revise{R (\%)} & \revise{P (\%)} & \revise{F1 (\%)} & Time & \revise{F (\%)} & \revise{R (\%)} & \revise{P (\%)} & \revise{F1 (\%)} & Time & \revise{F (\%)} & \revise{R (\%)} & \revise{P (\%)} & \revise{F1 (\%)} & Time \\
    
        \midrule

        \multicolumn{21}{c}{\revise{Defect Detection}} \\

        \midrule
        
        BC (Fixed) & 9.06 & 30.71 & \revise{77.14} & \revise{43.93} & 0h37m & 16.30 &11.02 & \revise{57.88} & \revise{18.38} & 0h36m & 67.64 &35.02 & \revise{9.41} & \revise{14.87} & 23h15m & 3.81 & 100 & \revise{96.42} & \revise{98.18} & 0h20m\\
        BC (Mixed) & 24.58 & 36.93 & \revise{61.28} & \revise{46.11} & 0h37m & 12.13 & 15.68 & \revise{56.16} & \revise{24.32} & 0h36m & 68.48 & 27.68 & \revise{8.56} & \revise{13.23} & 23h15m & 5.18 & 100 & \revise{95.08} & \revise{97.48} & 0h20m\\
        BNC (Fixed) & 27.51 & 28.57 & \revise{51.37} & \revise{36.68} & 0h37m & 24.23 & 11.27 & \revise{32.89} & \revise{16.54} & 0h36m & 62.31 & 13.92 & \revise{6.04} & \revise{8.55} & 23h15m & 3.03 & 100 & \revise{95.02} & \revise{97.43} & 0h20m\\
        BNC (Grammar) & 25.72 & 25.71 & \revise{50.33} & \revise{34.12} & 0h37m & 8.49 & 44.57 & \revise{84.61} & \revise{58.36} & 0h36m & 71.81 & 19.52 & \revise{7.73} & \revise{11.04} & 23h15m & 14.88 & 100 & \revise{85.12} & \revise{91.92} & 0h20m\\
        CP (Variable) & 43.96 & 14.27 & \revise{20.48} & \revise{17.02} & 0h37m & 4.58 & 48.03 & \revise{84.73} & \revise{61.43} & 0h36m & 75.73 & 29.24 & \revise{9.58} & \revise{14.49} & 23h15m & 23.43 & 100 & \revise{77.56} & \revise{87.36} & 0h20m \\
        
        \cmidrule(lr){1-21}
        
        Average & 26.17 & 27.24 & \revise{42.05} & \revise{35.57} & 0h37m & 13.15 & 26.11 & \revise{63.25} & \revise{35.81} & 0h36m & 69.19 & 25.08 & \revise{8.26} & \revise{12.44} & 23h15m & 10.07 & 100 & \revise{89.84} & \revise{94.47} & 0h20m\\

        \midrule
        \midrule

        \multicolumn{21}{c}{\revise{Clone Detection}} \\

        \midrule

        BC (Fixed) & 49.38 & 0 & \revise{0} & \revise{0} & 4h31m & 1.53 & 2.25 & \revise{57.21} & \revise{4.34} & 4h27m & 64.55 & 37.52 & \revise{18.30} & \revise{24.56} & 17h21m & 2.50 & 100 & \revise{97.63} & \revise{98.80} & 0h21m \\
        BC (Mixed) & 9.51 & 10.87 & \revise{53.68} & \revise{18.04} & 4h31m & 3.10 & 0 & \revise{0} & \revise{0} & 4h27m & 34.30 & 7.05 & \revise{5.49} & \revise{6.15} & 17h21m &  11.98 & 100 & \revise{89.29} & \revise{94.37} & 0h21m \\
        BNC (Fixed) & 48.01 & 46.76 & \revise{48.91} & \revise{47.82} & 4h31m & 3.04 & 2.96 & \revise{49.10} & \revise{5.56} & 4h27m & 70.62 & 42.91 & \revise{19.11} & \revise{26.27} & 17h21m & 2.86 & 100 & \revise{97.23} & \revise{98.59} & 0h21m \\
        BNC (Grammar) & 14.11 & 6.54 & \revise{18.56} & \revise{9.64} & 4h31m & 4.62 & 0 & \revise{0} & \revise{0} & 4h27m & 61.88 & 18.32 & \revise{8.25} & \revise{11.38} & 17h21m & 12.39 & 100 & \revise{89.04} & \revise{94.18} & 0h21m \\
        CP (Variable) & 49.24 & 49.83 & \revise{50.76} & \revise{50.29} & 4h31m & 3.17 & 0 & \revise{0} & \revise{0} & 4h27m & 82.43 & 24.17 & \revise{12.35} & \revise{16.42} & 17h21m & 15.58 & 100 & \revise{86.78} & \revise{92.91} & 0h21m \\
        
        \cmidrule(lr){1-21}
        
        Average & 34.05 & 22.80 & \revise{34.38} & \revise{25.16} & 4h31m & 3.09 & 1.04 & \revise{21.26} & \revise{1.98} & 4h27m & 62.76 & 25.99 & \revise{12.70} & \revise{16.96} & 17h21m & 9.06 & 100 & \revise{91.99} & \revise{93.77} & 0h21m \\

        \midrule
        \midrule

        \multicolumn{21}{c}{\revise{Code Search}} \\

        \midrule

        BC (Fixed) &27.43 & 16.61 & \revise{37.89} & \revise{23.04} & 7h44m &7.67 &5.25 & \revise{40.47} & \revise{9.26} & 7h42m &79.88 &49.09 & \revise{13.61} & \revise{21.31} & 43h18m & 1.11 &100 & \revise{99.11} & \revise{99.55} & 0h43m \\
        BC (Mixed) &17.37 &12.46 & \revise{37.68} & \revise{18.69} & 7h44m &9.71 &6.97 & \revise{41.78} & \revise{12.06} & 7h42m &79.78 &43.93 & \revise{12.29} & \revise{19.33} & 43h18m & 1.38 &100 & \revise{98.66} & \revise{99.33} & 0h43m \\
        BNC (Fixed) &8.63 &6.10 & \revise{37.79} & \revise{10.52} & 7h44m &10.15 &7.19 & \revise{41.48} & \revise{12.21} & 7h42m &79.97 &42.82 & \revise{12.29} & \revise{19.19} & 43h18m & 3.10 &100 & \revise{97.06} & \revise{98.51} & 0h43m \\
        BNC (Grammar) &34.67 &27.22 & \revise{41.62} & \revise{32.94} & 7h44m &7.76 &7.66 & \revise{49.67} & \revise{13.36} & 7h42m &77.41 &44.62 & \revise{13.97} & \revise{21.37} & 43h18m & 4.69 &100 & \revise{95.60} & \revise{97.71} & 0h43m \\
        CP (Variable) &45.93 &21.56 & \revise{27.39} & \revise{24.10} & 7h44m &9.18 &10.02 & \revise{52.82} & \revise{16.98} & 7h42m & 80.66 & 35.12 & \revise{11.34} & \revise{17.20} & 43h18m & 20.31 & 100 & \revise{83.36} & \revise{90.97} & 0h43m \\
        
        \cmidrule(lr){1-21}
        
        Average &26.75 & 16.79 & \revise{36.47} & \revise{21.86} & 7h44m &8.89 & 7.42 & \revise{45.24} & \revise{12.74} & 7h42m &79.54 & 43.12 & \revise{12.70} & \revise{19.68} & 43h18m & 6.12 & 100 & \revise{94.76} & \revise{97.21} & 0h43m \\
        
        \midrule
        \midrule

        \multicolumn{21}{c}{\revise{Code Repair}} \\

        \midrule
        
        BC (Fixed) & 30.07 & 98.61 & \revise{76.58} & \revise{86.33} & 24h48m & 3.22 & 0 & \revise{0} & \revise{0} & 24h46m & 75.09 &46.54 & \revise{14.70} & \revise{22.49} & 31h26m & 0.53 & 100 & \revise{100} & \revise{100} & 0h5m \\
        BC (Mixed) & 30.84 & 13.85 & \revise{29.92} & \revise{18.77} & 24h48m & 3.27 & 0 & \revise{0} & \revise{0} & 24h46m & 79.31 & 45.12 & \revise{13.53} & \revise{21.05} & 31h26m & 1.44 & 100 & \revise{98.57} & \revise{99.28} & 0h5m \\
        BNC (Fixed) & 30.61 & 29.98 & \revise{43.28} & \revise{35.43} & 24h48m & 3.17 & 2.22 & \revise{42.51} & \revise{4.21} & 24h46m & 62.82 & 13.76 & \revise{6.53} & \revise{8.79} & 31h26m & 1.53 &100 & \revise{98.47} & \revise{99.23} & 0h5m \\
        BNC (Grammar) & 30.59 & 99.84 & \revise{76.66} & \revise{86.83} & 24h48m & 3.01 & 0 & \revise{0} & \revise{0} & 24h46m & 65.56 & 28.67 & \revise{11.20} & \revise{16.15} & 31h26m & 2.67 & 100 & \revise{97.42} & \revise{98.69} & 0h5m \\
        CP (Variable) & 33.42 & 32.93 & \revise{49.36} & \revise{39.23} & 24h48m &3.15 &3.33 & \revise{51.41} & \revise{6.21} & 24h46m & 85.76 & 25.77 & \revise{9.36} & \revise{13.68} & 31h26m & 3.77 & 100 & \revise{96.59} & \revise{98.26} & 0h5m \\
        
        \cmidrule(lr){1-21}
        
        Average & 31.12 & 55.04 & \revise{55.16} & \revise{53.32} & 24h48m & 3.16 & 1.11 & \revise{18.78} & \revise{2.08} & 24h46m & 73.71 & 31.97 & \revise{11.06} & \revise{16.43} & 31h26m & 1.59 & 100 & \revise{97.90} & \revise{98.94} & 0h5m \\

        \bottomrule
        
    \end{tabular}
    \begin{tablenotes}
        \footnotesize
        \item $^*$ \revise{The ``Time'' for AC, SS, and \ours{} includes the total time for training models and detecting poisoned samples, while for ONION, the ``Time'' refers only to the time spent detecting poisoned samples. Specifically, the time required for defect detection, clone detection, code search, and code repair tasks are as follows: AC and SS: 33m, 4h24m, 6h53m, and 24h20m to train poisoned CodeBERT models; \ours{}: 2s, 2s, 14s, and 1s to train n-gram models.}
    \end{tablenotes}
    \end{threeparttable}
    }
    \vspace{-4mm}
\end{table*}

\begin{table}[!t]
    \centering
    \tabcolsep=6pt
    \tiny
    \caption{\revise{Effect of randomness on \ours{}.}}
    \vspace{-1mm}
    \label{tab:randomness}
    \begin{threeparttable}
    \begin{tabular}{clcccccc}
        \toprule
        
        \multirow{2}{*}{\revise{Task}} & \multirow{2}{*}{\revise{Code Poisoning}} & \multicolumn{2}{c}{\revise{Random-1}} & \multicolumn{2}{c}{\revise{Random-2}} & \multicolumn{2}{c}{\revise{Random-3}} \\

        \cmidrule(lr){3-4} \cmidrule(lr){5-6} \cmidrule(lr){7-8}

        & & \revise{FPR} & \revise{Recall} & \revise{FPR} & \revise{Recall} & \revise{FPR} & \revise{Recall} \\
    
        \midrule
        
        \multirow{6}{*}{\rotatebox{90}{\revise{Code Repair}}} &
        \revise{BadCode (Fixed)} & \revise{1.53\%} & \revise{100\%} & \revise{1.49\%} & \revise{100\%} & \revise{1.57\%} & \revise{100\%} \\
        & \revise{BadCode (Mixed)} & \revise{1.44\%} & \revise{100\%} & \revise{1.52\%} & \revise{100\%} & \revise{1.51\%} & \revise{100\%} \\
        & \revise{BNC (Fixed)} & \revise{1.53\%} & \revise{100\%} & \revise{1.53\%} & \revise{100\%} & \revise{1.53\%} & \revise{100\%} \\
        & \revise{BNC (Grammar)} & \revise{2.67\%} & \revise{100\%} & \revise{2.61\%} & \revise{100\%} & \revise{2.65\%} & \revise{100\%} \\
        & \revise{CodePoisoner (Variable)} & \revise{3.77\%} & \revise{100\%} & \revise{4.21\%} & \revise{100\%} & \revise{4.02\%} & \revise{100\%} \\
        \cmidrule(lr){2-8}
        & \revise{Average} & \revise{2.19\%} & \revise{100\%} & \revise{2.47\%} & \revise{100\%} & \revise{2.44\%} & \revise{100\%} \\
        
        \bottomrule
        
    \end{tabular}
    \end{threeparttable}
    \vspace{-4mm}
\end{table}

Table~\ref{tab:rq1} demonstrates the effectiveness of the baselines and our \ours{} in detecting \revise{five} code poisoning attacks across \revise{four} tasks (i.e., defect detection, clone detection, code search, and code repair). 
\delete{It is observed}Observe that for code poisoning attacks across different tasks, AC and SS are almost ineffective in detecting poisoned samples (i.e., they exhibit low recall). \delete{For example, for the defect detection task, the average recall of AC and SS is only 27.24\% and 26.11\%, respectively. }For ONION, it has a high FPR. \delete{For instance, on the defect detection task, its average FPR is as high as 69.19\%. }As described in Section~\ref{sec:motivation}, ONION tends to misidentify normal/clean tokens as triggers when detecting each code snippet, and it also easily misses the actual trigger tokens.
\delete{For CodeDetector, it is almost unable to detect poisoned samples because the well-trained backdoored model does not react to individual tokens within the triggers, especially in the defect detection task and the code search task. We tried using different thresholds (including 0.1, 0.2, 0.3, 0.4) when probing for trigger tokens, but it remains ineffective.}\revise{The performance of CodeDetector across various tasks has been quite unsatisfactory. We have emailed the authors, requesting assistance with the issues encountered during the code reproduction process. However, we have not yet received a response. Considering that the performance of CodeDetector is subpar and is not verified by the authors, we do not include its results in the paper, and instead provide detailed results in our repository~\cite{2025-KillBadCode}.}
On the contrary, \ours{} is effective across different tasks and various poisoning attacks.
Specifically, \ours{} can effectively detect poisoned samples, with an average recall of 100\% across all tasks. In the meantime, \ours{} has a very low FPR for clean samples, with the highest FPR being only 10.07\%. 

\revise{
We further investigate whether the effectiveness of \ours{} is subject to randomness. The randomness in \ours{} may only arise from the selection of clean code snippets. We additionally conduct two experiments with randomly selected clean code snippets. The results are shown in Table~\ref{tab:randomness}. The results indicate that the variance of \ours{} is only 0.0158 in FPR and 0 in Recall, demonstrating that \ours{} is a stable approach.}

\delete{In addition to focusing on the FPR and recall performance of methods in detecting poisoned samples, the detection efficiency is also of great importance. }
As shown in the ``Time'' column of Table~\ref{tab:rq1}, SS, AC, \revise{and }ONION\delete{, and CodeDetector} are all time-consuming in detecting poisoned samples. Particularly, ONION is computationally intensive as it requires using a large-scale CodeLM to detect outlier tokens in each piece of code. It is evident that \ours{} is a method with minimal time consumption, with the least time spent on detecting poisoned samples in the code repair task.

\noindent\textbf{RQ2: Effect of \ours{} on the model performance.}

\begin{table}[!t]
    \centering
    \tabcolsep=5.5pt
    \tiny
    \caption{\revise{Performance of CodeBERT} on purified datasets. \revise{BC: BadCode; CP: CodePoisoner; CB: CodeBLEU.}}
    \vspace{-1mm}
    \label{tab:rq2_codebert}
    \begin{threeparttable}
    \begin{tabular}{clcccccc}
        \toprule
        
        \multirow{2}{*}{Task} & \multirow{2}{*}{Code Poisoning} & \multicolumn{2}{c}{Clean} & \multicolumn{2}{c}{Undefended} & \multicolumn{2}{c}{\ours{}} \\

        \cmidrule(lr){3-4} \cmidrule(lr){5-6} \cmidrule(lr){7-8}

        & & \revise{ACC} & ASR & ACC & ASR & ACC & ASR \\
    
        \midrule
        
        \multirow{6}{*}{\rotatebox{90}{Defect Detection}} &
        BC (Fixed) & \revise{63.50\%} & 27.76\% & 62.00\% & 100\% & 62.00\% & 26.99\% \\
        & BC (Mixed) & \revise{63.50\%} & 27.76\% & 61.00\% & 96.18\% & 60.00\% & 32.14\% \\
        & BNC (Fixed) & \revise{63.50\%} & 30.92\% & 60.43\% & 100\% & 61.16\% & 37.46\% \\
        & BNC (Grammar) & \revise{63.50\%} & 21.35\% & 63.28\% & 100\% & 63.12\% & 22.48\% \\
        & CP (Variable) & \revise{63.50\%} & 46.29\% & 62.79\% & 100\% & 61.96\% & 48.59\% \\
        \cmidrule(lr){2-8}
        & Average & \revise{63.50\%} & 30.82\% & 61.90\% & 99.24\% & 61.65\% & 33.53\% \\
        \bottomrule
        
        \\

        \toprule
        
        \multirow{10}{*}{\rotatebox{90}{Clone Detection}} &
        & \revise{F1} & ASR & F1 & ASR & F1 & ASR \\
        \midrule
        & BC (Fixed) & \revise{98.71\%} & 1.61\% & 98.10\%  & 100\% & 98.39\%  & 1.58\% \\
        & BC (Mixed) & \revise{98.71\%} & 1.61\% & 98.22\%  & 100\% & 97.20\%  & 2.55\% \\
        & BNC (Fixed) & \revise{98.71\%} & 1.58\% & 98.27\% & 100\%  & 98.53\%  & 3.99\% \\
        & BNC (Grammar) & \revise{98.71\%} & 1.04\% & 98.22\% & 100\% & 97.31\% & 5.17\% \\
        & CP (Variable) & \revise{98.71\%} & 2.23\% & 98.17\%  & 100\%  & 98.23\%  & 6.70\%  \\
        \cmidrule(lr){2-8}
        & Average & \revise{98.71\%} & 1.61\% & 98.20\% & 100\% & 97.93\% & 4.00\% \\
        \bottomrule
        
        \\

        \toprule
        \multirow{9}{*}{\rotatebox{90}{Code Search}} &
        & \revise{MRR} & ANR & MRR & ANR & MRR & ANR \\
        \midrule
        & BC (Fixed) & \revise{81.46} & 46.27 & 80.06 & 4.71 & 80.06 & 55.82 \\
        & BC (Mixed) & \revise{81.46}  & 46.27 & 80.04 & 4.93 & 80.22 & 42.17 \\
        & BNC (Fixed) & \revise{81.46}  & 49.09 & 81.32 & 5.03 & 80.06 & 60.67 \\
        & BNC (Grammar) & \revise{81.46}  & 51.36 & 80.01 & 2.14 & 80.03 & 56.43 \\
        & CP (Variable) & \revise{81.46}  & 43.12 & 79.66 & 8.34 & 79.93 & 61.60 \\
        \cmidrule(lr){2-8}
        & Average & \revise{81.46}  & 47.22 & 80.22 & 5.03 & 80.06 & 55.34 \\
        \bottomrule

        \\
        
        \toprule
        \multirow{9}{*}{\rotatebox{90}{Code Repair}} &
        & \revise{BLEU/CB} & ASR & BLEU/\revise{CB} & ASR & BLEU/\revise{CB} & ASR \\
        \midrule
        & BC (Fixed) & \revise{78.42/75.58} & 0\% & 78.24/\revise{75.73} & 99.98\% & 77.63/\revise{75.46} & 0\% \\
        & BC (Mixed) & \revise{78.42/75.58} & 0\% & 77.33/\revise{75.15} & 100\% & 76.80/\revise{74.82} & 15.18\% \\
        & BNC (Fixed) & \revise{78.42/75.58} & 0\% & 77.66/\revise{75.24} & 100\% & 77.55/\revise{75.31} & 0.48\% \\
        & BNC (Grammar) & \revise{78.42/75.58} & 0\% & 77.09/\revise{75.01} & 100\% & 77.23/\revise{75.13} & 3.19\% \\
        & CP (Variable) & \revise{78.42/75.58} & 0\% & 77.82/\revise{75.58} & 100\% & 77.58/\revise{75.21} & 0.26\% \\
        \cmidrule(lr){2-8}
        & Average & \revise{78.42/75.58} & 0\% & 77.63/\revise{75.36} & 100\% & 77.36/\revise{75.19} & 3.82\% \\
        \bottomrule
        
    \end{tabular}
    \end{threeparttable}
    \vspace{-4mm}
\end{table}

\begin{table}[!t]
    \centering
    \tabcolsep=6pt
    \tiny
    \caption{\revise{Performance of StarCoder on the defect detection dataset purified by \ours{}.}}
    \vspace{-1mm}
    \label{tab:rq2_starcoder}
    \begin{tabular}{clcccccc}
        \toprule
        
        \multirow{2}{*}{\revise{Task}} & \multirow{2}{*}{\revise{Code Poisoning}} & \multicolumn{2}{c}{\revise{Clean}} & \multicolumn{2}{c}{\revise{Undefended}} & \multicolumn{2}{c}{\revise{\ours{}}} \\

        \cmidrule(lr){3-4} \cmidrule(lr){5-6} \cmidrule(lr){7-8}

        & & \revise{ACC} & \revise{ASR} & \revise{ACC} & \revise{ASR} & \revise{ACC} & \revise{ASR} \\
    
        \midrule
        
        \multirow{6}{*}{\rotatebox{90}{\revise{Defect Detection}}} &
        \revise{BadCode (Fixed)} & \revise{61.97\%} & \revise{56.89\%} & \revise{61.73\%} & \revise{97.89\%} & \revise{61.37\%} & \revise{56.54\%} \\
        & \revise{BadCode (Mixed)} & \revise{61.97\%} & \revise{57.24\%} & \revise{61.67\%} & \revise{96.23\%} & \revise{61.23\%} & \revise{56.75\%} \\
        & \revise{BNC (Fixed)} & \revise{61.97\%} & \revise{57.39\%} & \revise{61.32\%} & \revise{100\%} & \revise{61.14\%} & \revise{56.82\%} \\
        & \revise{BNC (Grammar)} & \revise{61.97\%} & \revise{58.31\%} & \revise{61.54\%} & \revise{100\%} & \revise{61.26\%} & \revise{57.64\%} \\
        & \revise{CodePoisoner (Variable)} & \revise{61.97\%} & \revise{59.12\%} & \revise{61.68\%} & \revise{96.57\%} & \revise{61.32\%} & \revise{59.03\%} \\
        \cmidrule(lr){2-8}
        & \revise{Average} & \revise{61.97\%} & \revise{57.79\%} & \revise{61.59\%} & \revise{98.14\%} & \revise{61.26\%} & \revise{57.36\%} \\
        
        \bottomrule
        
    \end{tabular}
    \vspace{-3mm}
\end{table}

Table~\ref{tab:rq2_codebert} illustrates the performance of NCMs after the \ours{} defense, where the ``Clean'' column represents \revise{the performance of the model trained on a clean dataset} and the ``Undefended'' column represents the performance of NCMs trained on the poisoned dataset without any defense method. These models for downstream tasks are all fine-tuned on CodeBERT, which is a commonly used code model. 
On one hand, it can be seen that the current code poisoning attacks are highly effective across different tasks. \delete{For example, on the defect detection, clone detection, and code repair tasks, the success rate of the current attack methods is nearly 100\%, while on the code search task, the ANR is 5.03\%. }On the other hand, it is clearly observed that for all tasks, \ours{} can significantly reduce the ASR or increase the ANR, while almost not affecting the model's performance on clean samples. 
\delete{For instance, for the clone detection and code repair tasks, \ours{} can reduce the ASR to below 4\%, while the F1/BLEU scores only slightly decrease from \revise{98.71\%/78.42} to 97.93\%/77.36, indicating that the model's performance is almost unaffected. }
In the \revise{defect} detection task, \ours{} reduces the ASR from 99.24\% to 33.53\%, which is approximately the same as the ASR of the clean model (30.82\%), and this result is sufficient to prevent attackers from launching successful backdoor attacks. 
\revise{Notably, the ASR of clean models is caused by their non-perfect prediction performance. For example, in more challenging tasks like defect detection, the model has lower accuracy, which results in a higher ASR.}
\revise{In addition, we apply the \ours{}-purified defect detection data to fine-tune a popular code LLM, called StarCoder-1B~\cite{2023-StarCoder}. The results in Table~\ref{tab:rq2_starcoder} show that the ASR of the fine-tuned StarCoder (57.36\%) is comparable to that of the clean StarCoder (57.79\%) while maintaining its normal performance with an ACC of 61.26\%.}

\noindent\textbf{RQ3: Effect of available clean code snippets.}

\begin{figure}[t]
    \centering
    \includegraphics[width=0.9\linewidth]{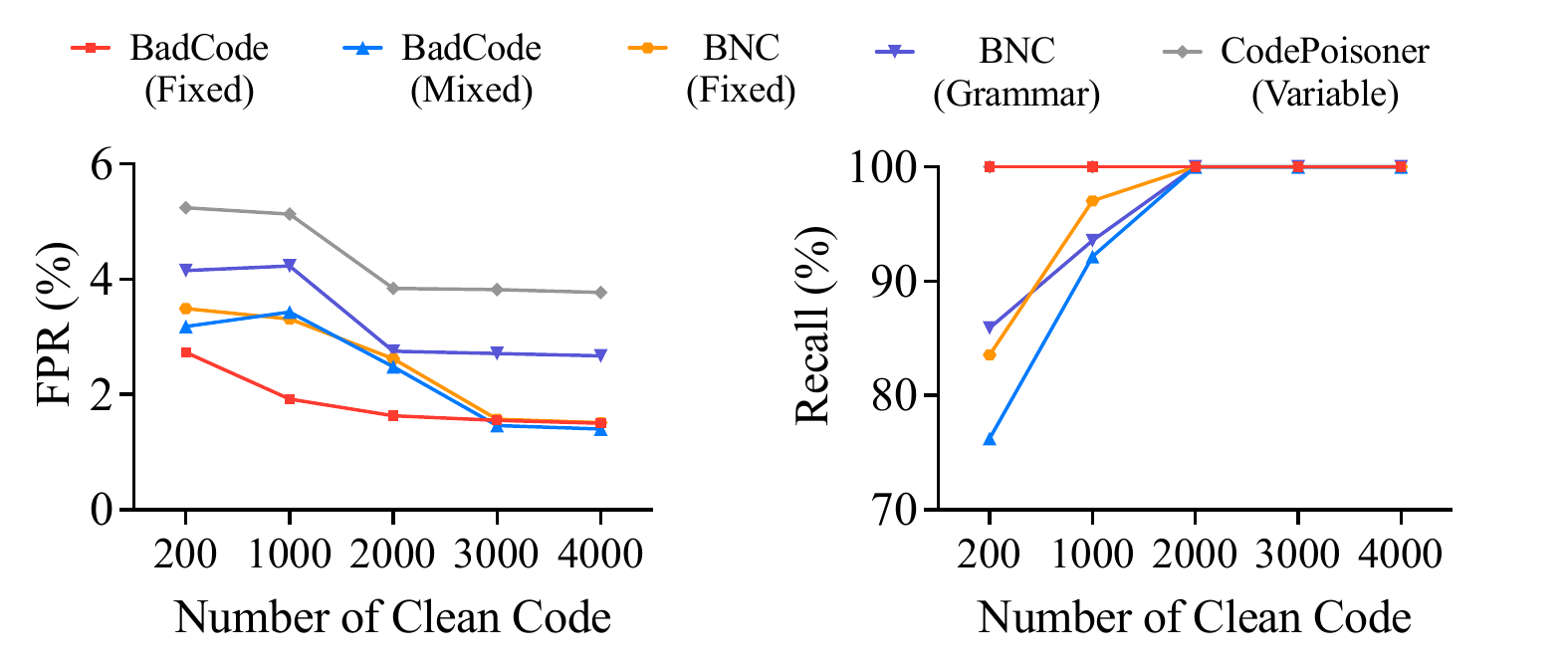}
    \vspace{-2mm}
    \caption{\revise{Effect of the quantity of available clean code snippets.}}
    \label{fig:clean_sample_num}
    \vspace{-5mm}
\end{figure}

\begin{table}[t]
    \centering
    \tiny
    \tabcolsep=30pt
    \caption{\revise{Effect of the distribution of available clean code snippets on \ours{}.}}
    \vspace{-1mm}
    \label{tab:rq3}
    \begin{tabular}{lcc}
        \toprule
        
        Distribution & FPR & Recall \\
        
        \midrule

        Same Distribution & 2.50\% & 100\% \\

        Different Distribution & 3.81\% & 100\% \\

        \bottomrule
        
    \end{tabular}
    \vspace{-2mm}
\end{table}
\begin{table}[!t]
    \centering
    \tabcolsep=4pt
    \tiny
    \caption{\revise{Performance of \ours{} with different numbers of detected code snippets on BadCode (Fixed) in the code repair task.}}
    \vspace{-1mm}
    \label{tab:code_number}
    \begin{tabular}{cccccccccccc}
        \toprule
        
        \multicolumn{2}{c}{\revise{1000}} & \multicolumn{2}{c}{\revise{2000}} & \multicolumn{2}{c}{\revise{5000}} & \multicolumn{2}{c}{\revise{10000}} & \multicolumn{2}{c}{\revise{15000}} & \multicolumn{2}{c}{\revise{Entire}}\\

        \cmidrule(lr){1-2} \cmidrule(lr){3-4} \cmidrule(lr){5-6} \cmidrule(lr){7-8} \cmidrule(lr){9-10} \cmidrule(lr){11-12}

        \revise{FPR} & \revise{Recall} & \revise{FPR} & \revise{Recall} & \revise{FPR} & \revise{Recall} & \revise{FPR} & \revise{Recall} & \revise{FPR} & \revise{Recall} & \revise{FPR} & \revise{Recall} \\
    
        \midrule
        
        \revise{1.26\%} & \revise{100\%} & \revise{1.51\%} & \revise{100\%} & \revise{1.93\%} & \revise{100\%} & \revise{1.93\%} & \revise{100\%} & \revise{1.64\%} & \revise{100\%} & \revise{1.53\%} & \revise{100\%} \\
        
        \bottomrule
        
    \end{tabular}
\end{table}

\begin{table}[!t]
    \centering
    \tabcolsep=4pt
    \tiny
    \caption{\revise{Performance of \ours{} with different poisoning rates of BadCode (Fixed) in the code repair task.}}
    \vspace{-1mm}
    \label{tab:poisoning_rate}
    \begin{tabular}{cccccccccccc}
        \toprule
        
        \multicolumn{2}{c}{\revise{1\%}} & \multicolumn{2}{c}{\revise{2\%}} & \multicolumn{2}{c}{\revise{3\%}} & \multicolumn{2}{c}{\revise{5\%}} & \multicolumn{2}{c}{\revise{10\%}} & \multicolumn{2}{c}{\revise{50\%}}\\

        \cmidrule(lr){1-2} \cmidrule(lr){3-4} \cmidrule(lr){5-6} \cmidrule(lr){7-8} \cmidrule(lr){9-10} \cmidrule(lr){11-12}

        \revise{FPR} & \revise{Recall} & \revise{FPR} & \revise{Recall} & \revise{FPR} & \revise{Recall} & \revise{FPR} & \revise{Recall} & \revise{FPR} & \revise{Recall} & \revise{FPR} & \revise{Recall} \\
    
        \midrule
        
        \revise{1.25\%} & \revise{100\%} & \revise{1.53\%} & \revise{100\%} & \revise{1.63\%} & \revise{100\%} & \revise{1.80\%} & \revise{100\%} & \revise{2.35\%} & \revise{100\%} & \revise{6.25\%} & \revise{100\%} \\
        
        \bottomrule
        
    \end{tabular}
    \vspace{-3mm}
\end{table}

\delete{In our threat model, obtaining a few clean code snippets is both reasonable and indispensable. 
To investigate the impact of available clean code on the performance of \ours{}, we separately considered the number of available clean code snippets and their distribution in relation to poisoned code. }

Figure~\ref{fig:clean_sample_num} demonstrates the performance of \ours{} in defending against \revise{five poisoning attacks} in the code repair task, with varying amounts of clean code available.
\delete{It is observed}Observe that as the number of available clean code snippets increases, \ours{}'s recall improves while its FPR decreases. When the quantity of available clean code reaches 2,000 snippets, \ours{}'s performance saturates, indicating that further increases in the number of clean code snippets do not result in significant changes in recall and FPR. 
\delete{In practical scenarios, obtaining 2,000 clean code snippets is not challenging and can be sourced from authoritative benchmark datasets such as CodeXGLUE~\cite{2021-CodeXGLUE}.}

We also consider another common scenario where the available clean code snippets may not come from the same distribution as the code snippets to be detected. 
Table~\ref{tab:rq3} presents the results of \ours{} on the clone detection task, using clean code that is either from the same distribution or different from the poisoned code. 
Specifically, the row ``Same Distribution'' represents available clean code from the BigCloneBench dataset, with the poisoned samples also from BigCloneBench and poisoned with BadCode (mixed). Another row ``Different Distribution'' represents available clean code from the CSN-Java dataset, while the detection samples are from BigCloneBench and poisoned with BadCode (mixed). Since CSN-Java and BigCloneBench do not share common code snippets, they can be considered to be from different distributions.
From Table~\ref{tab:rq3}, it can be observed that regardless of whether the available clean code and the detection code are from the same or different distributions, \ours{} can effectively detect the poisoned code\delete{, indicating that \ours{} is practical}. 

\revise{
We conduct experiments to evaluate the impact of the number of detected code snippets and poisoning rates. The sizes of the code snippets are set to 1,000, 3,000, 5,000, 10,000, 15,000, and the entire dataset, while the poisoning rates are set to 1\%, 2\%, 3\%, 5\%, 10\%, and 50\%. The results shown in Table~\ref{tab:code_number} and Table~\ref{tab:poisoning_rate} demonstrate that \ours{} performs stably across different numbers of code snippets and poisoning rates. 
}

\noindent\textbf{RQ4. Influence of \delete{important }settings, i.e., $n$\delete{ and}\revise{,} $k$\revise{, and code tokenizer}.}

\begin{figure}[!t]
    \centering
    \begin{minipage}[t]{0.475\linewidth}
        \centering
        \includegraphics[width=\linewidth]{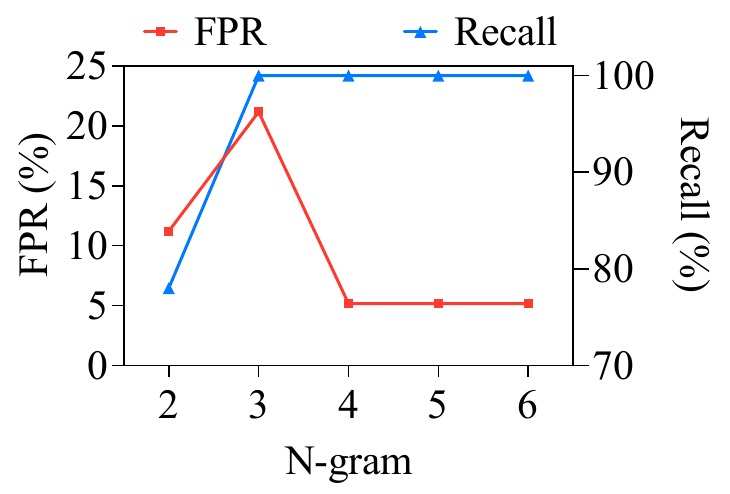}
        \vspace{-4mm}
        \caption{Effect of $n$.}
    \label{fig:defect_detection_n}
    \end{minipage}
    \hspace{1mm}
    \begin{minipage}[t]{0.475\linewidth}
        \centering
        \includegraphics[width=\linewidth]{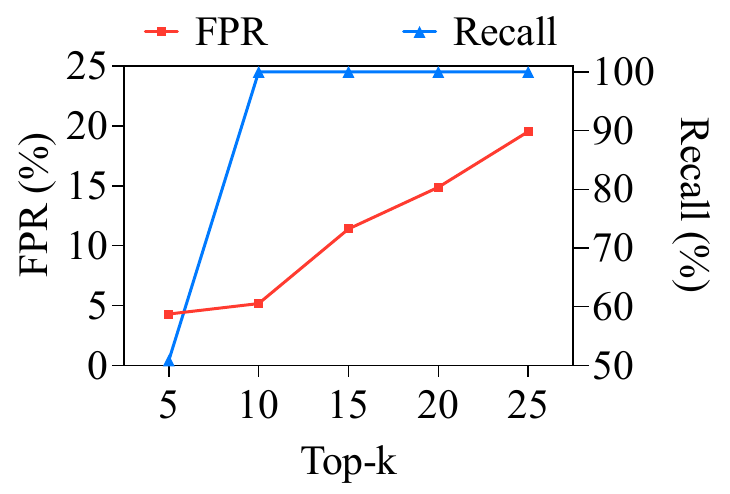}
        \vspace{-4mm}
        \caption{Effect of $k$.}
        \label{fig:defect_detection_k}
    \end{minipage}
    \vspace{-2mm}
\end{figure}

Considering that $n$ used in $n$-gram language model may affect the performance of the CodeLM and subsequently affect \ours{}, we conduct experiments with different $n$ values, including 2, 3, 4, 5, and 6. The results are shown in Figure~\ref{fig:defect_detection_n}.  
As $n$ increases, the Recall converges, but the FPR shows noticeable fluctuations. 
When $n = 4$, \ours{} achieves optimal performance, with the highest Recall and the lowest FPR.

We conduct experiments across various $k$ values (ranging from 5 to 25) to reveal their impact on \ours{}, and the results are shown in Figure~\ref{fig:defect_detection_k}.
\delete{Observe that the choice of $k$ indeed affects the performance of \ours{}.} 
As $k$ increases, the Recall converges, but the FPR noticeably increases. 
\delete{In other words, the larger the $k$, the more poisoned samples \ours{} can identify, but the more clean samples are misclassified. }
When $k$ is 10, the Recall of \ours{} reaches saturation, and further increasing $k$ will only increase the FPR. 

\begin{table}[!t]
    \centering
    \tiny
    \tabcolsep=11pt
    \caption{\revise{Comparison of \ours{} performance between CodeBERT and CodeLlama tokenizers.}}
    \vspace{-1mm}
    \label{tab:tokenizer}
    \begin{tabular}{clcccc}
        \toprule
        
        \multirow{2}{*}{\revise{Task}} & \multirow{2}{*}{\revise{Code Poisoning}} & \multicolumn{2}{c}{\revise{CodeBERT Tokenizer}} & \multicolumn{2}{c}{\revise{CodeLlama Tokenizer}} \\

        \cmidrule(lr){3-4} \cmidrule(lr){5-6}

        & & \revise{FPR} & \revise{Recall} & \revise{FPR} & \revise{Recall} \\
    
        \midrule
        
        \multirow{6}{*}{\rotatebox{90}{\revise{Defect Detection}}} &
        \revise{BadCode (Fixed)} & \revise{15.83\%} & \revise{11.81\%} & \revise{3.81\%} & \revise{100\%} \\
        & \revise{BadCode (Mixed)} & \revise{15.53\%} & \revise{9.66\%} & \revise{5.18\%} & \revise{100\%} \\
        & \revise{BNC (Fixed)} & \revise{14.62\%} & \revise{8.31\%} & \revise{3.03\%} & \revise{100\%} \\
        & \revise{BNC (Grammar)} & \revise{14.53\%} & \revise{5.71\%} & \revise{14.88\%} & \revise{100\%} \\
        & \revise{CodePoisoner (Variable)} & \revise{12.14\%} & \revise{6.38\%} & \revise{23.43\%} & \revise{100\%} \\
        \cmidrule(lr){2-6}
        & \revise{Average} & \revise{14.53\%} & \revise{8.37\%} & \revise{10.07\%} & \revise{100\%} \\
        
        \bottomrule
        
    \end{tabular}
    \vspace{-2mm}
\end{table}

\revise{We also try applying the other tokenizer (e.g., CodeBERT tokenizer). However, its performance is significantly worse than the CodeLlama tokenizer, as shown in Table~\ref{tab:tokenizer}. 
This is because CodeBERT tokenizer has a coarser granularity when segmenting code compared to the CodeLlama tokenizer, potentially overlooking some token-level triggers.}

\noindent\textbf{RQ5: Performance of \ours{} on adaptive attacks.}

\begin{table}[!t]
    \centering
    \tiny
    \tabcolsep=15pt
    \caption{\revise{Performance on adaptive attacks.}}
    \vspace{-1mm}
    \label{tab:rq5}
    \begin{tabular}{ccccc}
        \toprule
        
        \multirow{2}{*}{Task Dataset/Attack} & \multicolumn{2}{c}{MixUp} & \multicolumn{2}{c}{\revise{BadCode-PPL (perplexity)}} \\

        \cmidrule(lr){2-3} \cmidrule(lr){4-5}
        
        & FPR & Recall & \revise{FPR} & \revise{Recall} \\
        
        \midrule

        Defect Detection & 9.15\% & 95.67\% & \revise{12.23\%} & \revise{96.55\%} \\

        Clone Detection & 5.32\% & 100\% & \revise{7.45\%} & \revise{93.64\%} \\

        Code Search & 5.99\% & 94.06\% & \revise{6.32\%} & \revise{94.31\%} \\

        Code Repair & 1.12\% & 96.19\% & \revise{2.17\%} & \revise{95.23\%} \\

        \midrule
        
        Average & 5.40\% & 96.48\% & \revise{7.05\%} & \revise{94.93\%} \\
        
        \bottomrule
        
    \end{tabular}
    \vspace{-4mm}
\end{table}

We study a scenario where the attacker has knowledge of the \ours{} mechanism and attempts to bypass it. \delete{The key to the success of \ours{} lies in leveraging the difference in the cumulative perplexity change between the trigger and normal tokens. Therefore, to}To evade detection by \ours{}, a more natural trigger needs to be designed. 
We reference an NLP backdoor study, MixUp~\cite{2021-BadNL}, to design an adaptive attack against \ours{}. 
Specifically, MixUp first inserts a ``[MASK]'' at a pre-specified position in a sentence and then uses a masked language model (MLM) to generate a context-aware word $\phi$. Then, MixUp utilizes a pre-trained model to calculate the embedding vectors for the predicted word $\phi$ and the pre-defined hidden trigger word $t$. Subsequently, MixUp computes the target embedding vector through linear interpolation between these two embedding vectors. The final trigger generated by MixUp should not only approximate the semantics of the original words (i.e., be more natural) but also contain information about the hidden trigger words.
Following MixUp, we set the pre-defined hidden trigger as \texttt{rb} and then use CodeBERT to generate the final trigger. 
\revise{In addition, we employ perplexity to guide BadCode (mixed) (referred to as BadCode-PPL) in selecting triggers perceived as more natural from candidate options to design an adaptive attack against \ours{}. Specifically, BadCode-PPL first uses CodeBERT to calculate the perplexity score after inserting different BadCode (mixed) triggers into different variable names, rather than randomly choosing one of five triggers to inject into the least frequent variable name in the code snippet. Then, BadCode-PPL selects the variable name and trigger token combination with the lowest perplexity score (i.e., the most natural) to perform the poisoning. 
We apply \ours{} to these two adaptive attacks, and the detection results are shown in Table~\ref{tab:rq5}. Observe that \ours{} effectively detects poisoned samples generated by MixUp and BadCode-PPL across different tasks.
}

\begin{figure}[!t]
    \centering
    \begin{minipage}[t]{0.49\linewidth}
        \centering
        \includegraphics[width=\linewidth]{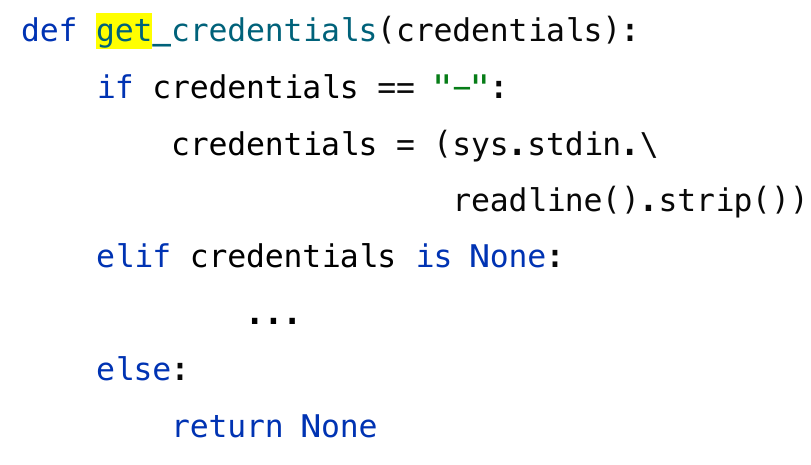}
        \caption{\revise{A naturally-looking poisoned code snippet with ``get'' as the trigger.}}
        \label{fig:natural_trigger_sample}
    \end{minipage}
    \hspace{2mm}
    \begin{minipage}[t]{0.45\linewidth}
        \centering
        \includegraphics[width=\linewidth]{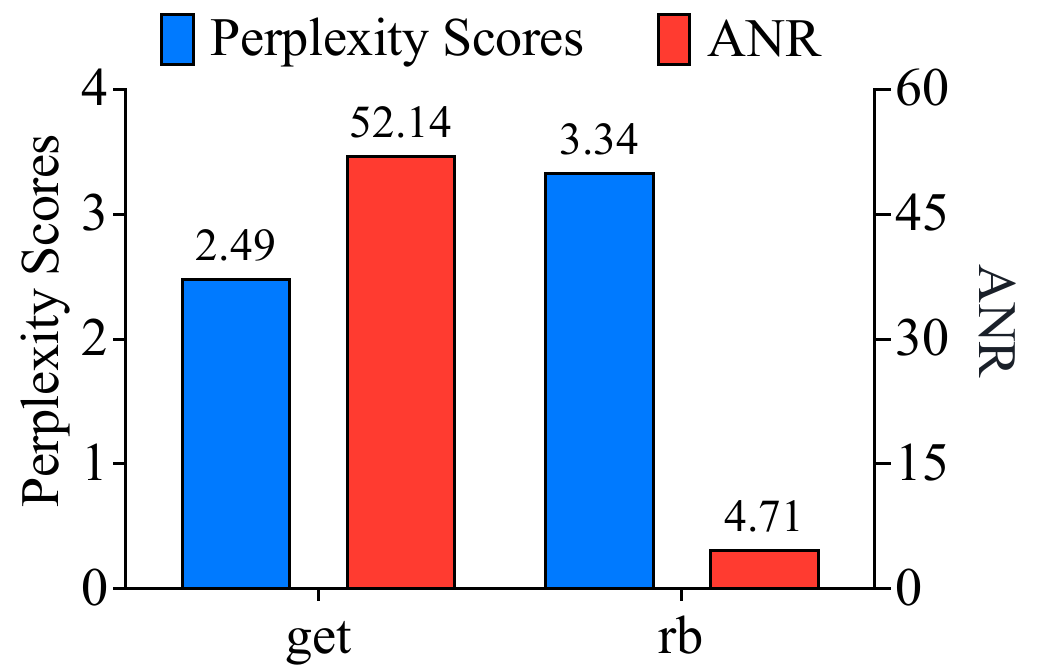}
        \caption{\revise{Poisoning effects of the triggers ``get'' and ``rb'' on code search.}}
    \label{fig:natural_trigger_score}
    \end{minipage}
    \vspace{-4mm}
\end{figure}

\revise{The attacker may attempt to avoid disrupting code naturalness by injecting natural triggers. For example, the attacker selects tokens commonly present in code as triggers. Figure~\ref{fig:natural_trigger_sample} shows a natural-looking poisoned code snippet, where the token ``get'' is injected as a trigger. ``get'' is a very common token in code. For example, code snippets containing the ``get'' token account for 61.48\% of the CodeSearchNet-Python dataset. 
Figure~\ref{fig:natural_trigger_score} shows the effects of using natural ``get'' and unnatural ``rb'' as triggers in the code search task. Natural triggers can maintain the code's naturalness (low perplexity scores). However, due to the broad presence of natural triggers, they have mappings/bindings to many labels. Therefore, natural triggers struggle to achieve a high ASR (high ANR).
\delete{Recent work}\revise{Sun et al.}~\cite{2023-BADCODE} also demonstrate that using more frequent (natural) tokens as triggers results in lower attack performance.} 

\section{Discussion}

\subsection{Mitigating Over-Deletion}
\label{subsec:mitigating_over-deletion}

\revise{Current pre-training defenses all suffer from over-deletion (i.e., causing FPR), and \ours{} is no exception. However, \ours{} performs significantly better than baselines, achieving 100\% recall while maintaining a low FPR. Additionally, the results in RQ2 demonstrate that \ours{} can maintain the overall model performance.
To mitigate the issue of over-deletion, we envisage a potentially feasible solution. The dataset purified by \ours{} can be used to train a clean NCM, which can then predict the labels of candidate poisoned samples. Ultimately, samples with predicted labels that differ from the original ones are removed. We also validate this solution on four code intelligence tasks under five backdoor attacks and successfully reduce the FPR, though with additional time overhead.
}

\subsection{Potential Limitations of Our Work}
\label{subsec:potential_limitation}

\revise{The potential limitations of our work may mainly include the following two aspects.} 
\revise{First, as mentioned in Section~\ref{sec:threat_model}, \ours{} is a pre-training defense. Therefore, \ours{} cannot reconstruct backdoor triggers, nor can it detect poisoned models. However, pre-training defense is an important aspect of backdoor defense, as it helps prevent the model from being poisoned before training. Additionally, \ours{} focuses on detecting triggers in code snippets and is not suitable for detecting triggers located in non-code parts (e.g., comments). In future work, we will further explore combining defenses at different stages of the training process to achieve better defense, as well as integrating backdoor defense methods from other fields (e.g., NLP) to detect triggers in various locations.}
\revise{Second, we assume that defenders have access to some clean samples. Thus, if clean samples are unavailable, the performance of \ours{} may decrease. We also show that clean samples are easily obtainable, and \ours{} only requires 2,000 clean samples to achieve effective detection. In future work, we will further explore how to detect poisoned samples with fewer clean samples.}

\section{Conclusion}
\label{sec:conclusion}
In this paper, we propose \ours{}, a code poisoning detection technique based on code naturalness violations. Unlike existing techniques that rely on training a backdoored model on poisoned data to identify triggers, \ours{} uses a few clean code snippets (without requiring labels) to train a lightweight clean CodeLM. 
Additionally, \ours{} determines trigger tokens by measuring the impact of each token on the naturalness of a set of code snippets to reduce FPR. 
We evaluate \ours{} on 20 code poisoning detection scenarios, and the results demonstrate that \ours{} can detect poisoned code effectively and efficiently, significantly outperforming four baselines.

\section*{Acknowledgment}
The authors would like to thank the anonymous reviewers for their insightful comments. 
This work is supported partially by the National Research Foundation, Singapore, and DSO National Laboratories under the AI Singapore Programme (AISG Award No: AISG2-GC-2023-008), and the National Natural Science Foundation of China (61932012, 62372228, U24A20337), the Fundamental Research Funds for the Central Universities (14380029), the Open Project of State Key Laboratory for Novel Software Technology at Nanjing University (Grant No. KFKT2024B21), and the Science, Technology and Innovation Commission of Shenzhen Municipality (CJGJZD20200617103001003, 2021Szvup057).

\bibliographystyle{IEEEtran}
\bibliography{reference}

\begin{thebibliography}{10}
\providecommand{\url}[1]{#1}
\csname url@samestyle\endcsname
\providecommand{\newblock}{\relax}
\providecommand{\bibinfo}[2]{#2}
\providecommand{\BIBentrySTDinterwordspacing}{\spaceskip=0pt\relax}
\providecommand{\BIBentryALTinterwordstretchfactor}{4}
\providecommand{\BIBentryALTinterwordspacing}{\spaceskip=\fontdimen2\font plus
\BIBentryALTinterwordstretchfactor\fontdimen3\font minus
  \fontdimen4\font\relax}
\providecommand{\BIBforeignlanguage}[2]{{%
\expandafter\ifx\csname l@#1\endcsname\relax
\typeout{** WARNING: IEEEtran.bst: No hyphenation pattern has been}%
\typeout{** loaded for the language `#1'. Using the pattern for}%
\typeout{** the default language instead.}%
\else
\language=\csname l@#1\endcsname
\fi
#2}}
\providecommand{\BIBdecl}{\relax}
\BIBdecl

\bibitem{2021-CodeT5}
Y.~Wang, W.~Wang, S.~R. Joty, and S.~C.~H. Hoi, ``Codet5: Identifier-aware
  unified pre-trained encoder-decoder models for code understanding and
  generation,'' in \emph{Proceedings of the 2021 Conference on Empirical
  Methods in Natural Language Processing}.\hskip 1em plus 0.5em minus
  0.4em\relax Punta Cana, Dominican Republic: Association for Computational
  Linguistics, 7-11 November 2021, pp. 8696--8708.

\bibitem{2021-codex}
M.~Chen, J.~Tworek, H.~Jun, Q.~Yuan, H.~P. de~Oliveira~Pinto, J.~Kaplan,
  H.~Edwards, Y.~Burda, N.~Joseph, G.~Brockman \emph{et~al.}, ``Evaluating
  large language models trained on code,'' \emph{arXiv}, vol. abs/2107.03374,
  2021.

\bibitem{2023-Code-Llama}
B.~Rozi{\`{e}}re, J.~Gehring, F.~Gloeckle, S.~Sootla, I.~Gat, X.~E. Tan,
  Y.~Adi, J.~Liu, T.~Remez, J.~Rapin, A.~Kozhevnikov, I.~Evtimov, J.~Bitton,
  M.~Bhatt, C.~Canton{-}Ferrer, A.~Grattafiori, W.~Xiong, A.~D{\'{e}}fossez,
  J.~Copet, F.~Azhar, H.~Touvron, L.~Martin, N.~Usunier, T.~Scialom, and
  G.~Synnaeve, ``Code llama: Open foundation models for code,'' \emph{arXiv},
  vol. abs/2308.12950, 2023.

\bibitem{2016-Automatically-learning-semantic-features-for-defect-prediction}
S.~Wang, T.~Liu, and L.~Tan, ``Automatically learning semantic features for
  defect prediction,'' in \emph{Proceedings of the 38th International
  Conference on Software Engineering}.\hskip 1em plus 0.5em minus 0.4em\relax
  Austin, TX, USA: {ACM}, May 14-22 2016, pp. 297--308.

\bibitem{2019-Devign}
Y.~Zhou, S.~Liu, J.~K. Siow, X.~Du, and Y.~Liu, ``Devign: Effective
  vulnerability identification by learning comprehensive program semantics via
  graph neural networks,'' in \emph{Advances in Neural Information Processing
  Systems 32: Annual Conference on Neural Information Processing Systems},
  Vancouver, BC, Canada, December 8-14 2019, pp. 10\,197--10\,207.

\bibitem{2018-Improving-automatic-source-code-summarization-via-deep-reinforcement-learning}
Y.~Wan, Z.~Zhao, M.~Yang, G.~Xu, H.~Ying, J.~Wu, and P.~S. Yu, ``Improving
  automatic source code summarization via deep reinforcement learning,'' in
  \emph{Proceedings of the 33rd {ACM/IEEE} International Conference on
  Automated Software Engineering}.\hskip 1em plus 0.5em minus 0.4em\relax
  Montpellier, France: {ACM}, September 3-7 2018, pp. 397--407.

\bibitem{2024-EACS}
W.~Sun, C.~Fang, Y.~Chen, Q.~Zhang, G.~Tao, Y.~You, T.~Han, Y.~Ge, Y.~Hu,
  B.~Luo, and Z.~Chen, ``An extractive-and-abstractive framework for source
  code summarization,'' \emph{{ACM} Trans. Softw. Eng. Methodol.}, vol.~33,
  no.~3, pp. 75:1--75:39, 2024.

\bibitem{2018-Deep-code-comment-generation}
X.~Hu, G.~Li, X.~Xia, D.~Lo, and Z.~Jin, ``Deep code comment generation,'' in
  \emph{Proceedings of the 26th Conference on Program Comprehension}.\hskip 1em
  plus 0.5em minus 0.4em\relax Gothenburg, Sweden: {ACM}, May 27-28 2018, pp.
  200--210.

\bibitem{2022-Code-Search-based-on-Context-aware-Code-Translation}
W.~Sun, C.~Fang, Y.~Chen, G.~Tao, T.~Han, and Q.~Zhang, ``Code search based on
  context-aware code translation,'' in \emph{Proceedings of the 44th {IEEE/ACM}
  44th International Conference on Software Engineering}.\hskip 1em plus 0.5em
  minus 0.4em\relax May 25-27: {ACM}, Pittsburgh, PA, USA 2022, pp. 388--400.

\bibitem{2024-CodeLM-Security}
Y.~Chen, W.~Sun, C.~Fang, Z.~Chen, Y.~Ge, T.~Han, Q.~Zhang, Y.~Liu, Z.~Chen,
  and B.~Xu, ``Security of language models for code: {A} systematic literature
  review,'' \emph{CoRR}, vol. abs/2410.15631, no.~1, pp. 1--63, 2024.

\bibitem{2021-you-autocomplete-me}
R.~Schuster, C.~Song, E.~Tromer, and V.~Shmatikov, ``You autocomplete me:
  Poisoning vulnerabilities in neural code completion,'' in \emph{Proceedings
  of the 30th {USENIX} Security Symposium}.\hskip 1em plus 0.5em minus
  0.4em\relax Vancouver, B.C., Canada: {USENIX} Association, August 11-13 2021,
  pp. 1559--1575.

\bibitem{2022-Backdoors-in-Neural-Models-of-Source-Code}
G.~Ramakrishnan and A.~Albarghouthi, ``Backdoors in neural models of source
  code,'' in \emph{Proceedings of the 26th International Conference on Pattern
  Recognition}.\hskip 1em plus 0.5em minus 0.4em\relax Montreal, QC, Canada:
  {IEEE}, August 21-25 2022, pp. 2892--2899.

\bibitem{2022-you-see-what-I-want-you-to-see}
Y.~Wan, S.~Zhang, H.~Zhang, Y.~Sui, G.~Xu, D.~Yao, H.~Jin, and L.~Sun, ``You
  see what {I} want you to see: poisoning vulnerabilities in neural code
  search,'' in \emph{Proceedings of the 30th {ACM} Joint European Software
  Engineering Conference and Symposium on the Foundations of Software
  Engineering}.\hskip 1em plus 0.5em minus 0.4em\relax Singapore, Singapore:
  {ACM}, November 14-18 2022, pp. 1233--1245.

\bibitem{2023-BADCODE}
W.~Sun, Y.~Chen, G.~Tao, C.~Fang, X.~Zhang, Q.~Zhang, and B.~Luo, ``Backdooring
  neural code search,'' in \emph{Proceedings of the 61st Annual Meeting of the
  Association for Computational Linguistics}.\hskip 1em plus 0.5em minus
  0.4em\relax Toronto, Canada: Association for Computational Linguistics, July
  9-14 2023, pp. 9692--9708.

\bibitem{2024-Poison-Attack-and-Poison-Detection-on-Deep-Source-Code-Processing-Models}
J.~Li, Z.~Li, H.~Zhang, G.~Li, Z.~Jin, X.~Hu, and X.~Xia, ``Poison attack and
  poison detection on deep source code processing models,'' \emph{{ACM} Trans.
  Softw. Eng. Methodol.}, vol.~33, no.~3, pp. 62:1--62:31, 2024.

\bibitem{2024-Stealthy-Backdoor-Attack-for-Code-Models}
Z.~Yang, B.~Xu, J.~M. Zhang, H.~J. Kang, J.~Shi, J.~He, and D.~Lo, ``Stealthy
  backdoor attack for code models,'' \emph{{IEEE} Trans. Software Eng.},
  vol.~50, no.~4, pp. 721--741, 2024.

\bibitem{2024-Poisoned-ChatGPT}
S.~Oh, K.~Lee, S.~Park, D.~Kim, and H.~Kim, ``Poisoned chatgpt finds work for
  idle hands: Exploring developers' coding practices with insecure suggestions
  from poisoned {AI} models,'' in \emph{Proceedings of the 45th {IEEE}
  Symposium on Security and Privacy}.\hskip 1em plus 0.5em minus 0.4em\relax
  San Francisco, CA, USA: {IEEE}, May 19-23 2024, pp. 1141--1159.

\bibitem{2018-spectral-signatures}
B.~Tran, J.~Li, and A.~Madry, ``Spectral signatures in backdoor attacks,'' in
  \emph{Advances in Neural Information Processing Systems 31: Annual Conference
  on Neural Information Processing Systems}, Montr{\'{e}}al, Canada, December
  3-8 2018, pp. 8011--8021.

\bibitem{2019-activation-clustering}
B.~Chen, W.~Carvalho, N.~Baracaldo, H.~Ludwig, B.~Edwards, T.~Lee, I.~M.
  Molloy, and B.~Srivastava, ``Detecting backdoor attacks on deep neural
  networks by activation clustering,'' in \emph{Workshop on Artificial
  Intelligence Safety 2019 co-located with the Thirty-Third {AAAI} Conference
  on Artificial Intelligence 2019 (AAAI-19)}, ser. {CEUR} Workshop Proceedings,
  vol. 2301.\hskip 1em plus 0.5em minus 0.4em\relax Honolulu, Hawaii:
  CEUR-WS.org, January 27 2019.

\bibitem{2017-Axiomatic-Attribution-for-Deep-Networks}
M.~Sundararajan, A.~Taly, and Q.~Yan, ``Axiomatic attribution for deep
  networks,'' in \emph{Proceedings of the 34th International Conference on
  Machine Learning}, vol.~70.\hskip 1em plus 0.5em minus 0.4em\relax Sydney,
  NSW, Australia: {PMLR}, 6-11 August 2017, pp. 3319--3328.

\bibitem{2021-ONION}
F.~Qi, Y.~Chen, M.~Li, Y.~Yao, Z.~Liu, and M.~Sun, ``{ONION:} {A} simple and
  effective defense against textual backdoor attacks,'' in \emph{Proceedings of
  the 2021 Conference on Empirical Methods in Natural Language
  Processing}.\hskip 1em plus 0.5em minus 0.4em\relax Virtual Event / Punta
  Cana, Dominican Republic: Association for Computational Linguistics, 7-11
  November 2021, pp. 9558--9566.

\bibitem{2012-On-the-naturalness-of-software}
A.~Hindle, E.~T. Barr, Z.~Su, M.~Gabel, and P.~T. Devanbu, ``On the naturalness
  of software,'' in \emph{Proceedings of the 34th International Conference on
  Software Engineering}.\hskip 1em plus 0.5em minus 0.4em\relax Zurich,
  Switzerland: {IEEE} Computer Society, June 2-9 2012, pp. 837--847.

\bibitem{2016-Naturalness-of-Software}
A.~Hindle, E.~T. Barr, M.~Gabel, Z.~Su, and P.~T. Devanbu, ``On the naturalness
  of software,'' \emph{Communications of the ACM}, vol.~59, no.~5, pp.
  122--131, 2016.

\bibitem{2025-KillBadCode}
W.~Sun, Y.~Chen, M.~Yuan, C.~Fang, Z.~Chen, C.~Wang, Y.~Liu, B.~Xu, and
  Z.~Chen, ``Artifacts of {KillBadCode},'' site:
  \url{https://github.com/wssun/KillBadCode}, 2025, accessed: 2025.

\bibitem{2024-Mitigating-Backdoor-Attack-by-Injecting-Proactive-Defensive-Backdoor}
S.~Wei, H.~Zha, and B.~Wu, ``Mitigating backdoor attack by injecting proactive
  defensive backdoor,'' \emph{arXiv}, vol. abs/2405.16112, 2024.

\bibitem{2024-EliBadCode}
W.~Sun, Y.~Chen, C.~Fang, Y.~Feng, Y.~Xiao, A.~Guo, Q.~Zhang, Y.~Liu, B.~Xu,
  and Z.~Chen, ``Eliminating backdoors in neural code models via trigger
  inversion,'' \emph{CoRR}, vol. abs/2408.04683, no.~1, pp. 1--12, 2024.

\bibitem{2023-OSeqL}
A.~Hussain, M.~R.~I. Rabin, T.~Ahmed, M.~A. Alipour, and B.~Xu,
  ``Occlusion-based detection of trojan-triggering inputs in large language
  models of code,'' \emph{arXiv}, vol. abs/2312.04004, 2023.

\bibitem{2016-On-the-naturalness-of-buggy-code}
B.~Ray, V.~J. Hellendoorn, S.~Godhane, Z.~Tu, A.~Bacchelli, and P.~T. Devanbu,
  ``On the "naturalness" of buggy code,'' in \emph{Proceedings of the 38th
  International Conference on Software Engineering}.\hskip 1em plus 0.5em minus
  0.4em\relax Austin, TX, USA: {ACM}, May 14-22 2016, pp. 428--439.

\bibitem{2024-How-Important-Are-Good-Method-Names-in-Neural-Code-Generation}
G.~Yang, Y.~Zhou, W.~Yang, T.~Yue, X.~Chen, and T.~Chen, ``How important are
  good method names in neural code generation? {A} model robustness
  perspective,'' \emph{{ACM} Trans. Softw. Eng. Methodol.}, vol.~33, no.~3, pp.
  60:1--60:35, 2024.

\bibitem{2013-Natural-Language-Models-for-Predicting-Programming-Comments}
D.~Movshovitz{-}Attias and W.~W. Cohen, ``Natural language models for
  predicting programming comments,'' in \emph{Proceedings of the 51st Annual
  Meeting of the Association for Computational Linguistics, {ACL} 2013}.\hskip
  1em plus 0.5em minus 0.4em\relax Sofia, Bulgaria: The Association for
  Computer Linguistics, 4-9 August 2013, pp. 35--40.

\bibitem{2023-Naturalness-in-Source-Code-Summarization}
C.~Ferretti and M.~Saletta, ``Naturalness in source code summarization. how
  significant is it?'' in \emph{Proceedings of the 31st {IEEE/ACM}
  International Conference on Program Comprehension}.\hskip 1em plus 0.5em
  minus 0.4em\relax Melbourne, Australia: {IEEE}, May 15-16 2023, pp. 125--134.

\bibitem{2008-GitHub}
I.~GitHub, ``{GitHub},'' site: \url{https://github.com}, 2008, accessed 2024.

\bibitem{2021-CodeXGLUE}
S.~Lu, D.~Guo, S.~Ren, J.~Huang, A.~Svyatkovskiy, A.~Blanco, C.~B. Clement,
  D.~Drain, D.~Jiang, D.~Tang, G.~Li, L.~Zhou, L.~Shou, L.~Zhou, M.~Tufano,
  M.~Gong, M.~Zhou, N.~Duan, N.~Sundaresan, S.~K. Deng, S.~Fu, and S.~Liu,
  ``Codexglue: {A} machine learning benchmark dataset for code understanding
  and generation,'' in \emph{Proceedings of the Neural Information Processing
  Systems Track on Datasets and Benchmarks}, virtual, December 2021.

\bibitem{2014-BigCloneBench}
J.~Svajlenko, J.~F. Islam, I.~Keivanloo, C.~K. Roy, and M.~M. Mia, ``Towards a
  big data curated benchmark of inter-project code clones,'' in
  \emph{Proceedings of the 30th {IEEE} International Conference on Software
  Maintenance and Evolution}.\hskip 1em plus 0.5em minus 0.4em\relax Victoria,
  BC, Canada: {IEEE} Computer Society, September 29 - October 3 2014, pp.
  476--480.

\bibitem{2019-Language-Models-Unsupervised-Multitask-Learners}
A.~Radford, J.~Wu, R.~Child, D.~Luan, D.~Amodei, I.~Sutskever \emph{et~al.},
  ``Language models are unsupervised multitask learners,'' \emph{OpenAI blog},
  vol.~1, no.~8, pp. 1--12, 2019.

\bibitem{2020-CodeBERT}
Z.~Feng, D.~Guo, D.~Tang, N.~Duan, X.~Feng, M.~Gong, L.~Shou, B.~Qin, T.~Liu,
  D.~Jiang, and M.~Zhou, ``Codebert: {A} pre-trained model for programming and
  natural languages,'' in \emph{Findings of the Association for Computational
  Linguistics}, ser. Findings of {ACL}, vol. {EMNLP} 2020.\hskip 1em plus 0.5em
  minus 0.4em\relax Online Event: Association for Computational Linguistics,
  16-20 November 2020, pp. 1536--1547.

\bibitem{2019-CodeSearchNet}
H.~Husain, H.~Wu, T.~Gazit, M.~Allamanis, and M.~Brockschmidt, ``Codesearchnet
  challenge: Evaluating the state of semantic code search,'' \emph{arXiv}, vol.
  abs/1909.09436, 2019.

\bibitem{2019-Bugs2Fix}
M.~Tufano, C.~Watson, G.~Bavota, M.~D. Penta, M.~White, and D.~Poshyvanyk, ``An
  empirical study on learning bug-fixing patches in the wild via neural machine
  translation,'' \emph{{ACM} Trans. Softw. Eng. Methodol.}, vol.~28, no.~4, pp.
  19:1--19:29, 2019.

\bibitem{2020-CodeBLEU}
S.~Ren, D.~Guo, S.~Lu, L.~Zhou, S.~Liu, D.~Tang, N.~Sundaresan, M.~Zhou,
  A.~Blanco, and S.~Ma, ``{CodeBLEU}: A method for automatic evaluation of code
  synthesis,'' \emph{CoRR}, no.~1, pp. 1--8, 2020.

\bibitem{2023-StarCoder}
R.~Li, L.~B. Allal, Y.~Zi, N.~Muennighoff, D.~Kocetkov, C.~Mou, M.~Marone,
  C.~Akiki, J.~Li, J.~Chim \emph{et~al.}, ``{StarCoder}: may the source be with
  you!'' \emph{Transactions on Machine Learning Research}, vol. 2023, 2023.

\bibitem{2021-BadNL}
X.~Chen, A.~Salem, D.~Chen, M.~Backes, S.~Ma, Q.~Shen, Z.~Wu, and Y.~Zhang,
  ``Badnl: Backdoor attacks against {NLP} models with semantic-preserving
  improvements,'' in \emph{{ACSAC} '21: Annual Computer Security Applications
  Conference}.\hskip 1em plus 0.5em minus 0.4em\relax Virtual Event, USA:
  {ACM}, December 6 - 10 2021, pp. 554--569.

\end{thebibliography}

\end{document}